\documentclass[hidelinks,aps,pra,twocolumn,superscriptaddress,10pt]{revtex4-2} 
\usepackage{graphicx}
\usepackage{color}
\usepackage{bbm}
\usepackage{amsfonts}
\usepackage{amssymb}
\usepackage{amsmath}
\usepackage{color}

\def\black#1{{\color{black} {#1}}}
\newcommand{\be}[1]{\begin{equation}\label{#1}}
\newcommand{\ee}{\end{equation}}
\newcommand{\bea}[1]{\begin{eqnarray}\label{#1}}
\newcommand{\eea}{\end{eqnarray}}

\newcommand{\Eq}[1]{Eq. \ref{#1}}     
\newcommand{\Sec}[1]{Section~\ref{#1}}

\newcommand{\bsub}{\begin{subequations}}
\newcommand{\esub}{\end{subequations}}

\DeclareMathOperator{\Tr}{Tr}

\usepackage{fancyhdr}
\begin{document}
\fancyhead[R]{\ifnum\value{page}<2\relax\else\thepage\fi}

\title{Canonical Quantum Coarse-Graining and Surfaces of Ignorance}
\author{Shannon Ray}
\affiliation{Information Directorate, Air Force Research Laboratory, Rome, NY 13441, USA}
\author{\black{Paul M.  Alsing}}
\affiliation{Information Directorate, Air Force Research Laboratory, Rome, NY 13441, USA}
\author{Carlo Cafaro}
\affiliation{SUNY Polytechnic Institute, 12203 Albany, New York, USA}
\author{Shelton Jacinto}
\affiliation{Information Directorate, Air Force Research Laboratory, Rome, NY 13441, USA}

\begin{abstract}  

In this paper we introduce a canonical quantum coarse-graining and use negentropy to connect ignorance as measured by quantum information entropy and ignorance related to quantum coarse-graining.  For our procedure, macro-states are the set of purifications $\{|\bar{\Gamma}^{\rho}\rangle\}$ associated with density operator $\rho$ and  micro-states are elements of $\{|\bar{\Gamma}^{\rho}\rangle\}$.  Unlike other quantum coarse-graining procedures, ours always gives a well-defined unique coarse-graining of phase space.  Our coarse-graining is also unique in that the volumes of phase space associated with macro-states are computed from differential manifolds whose metric components are constructed from the Lie group symmetries that generate $\{|\bar{\Gamma}^{\rho}\rangle\}$.  We call these manifolds surfaces of ignorance, and their volumes quantify the lack of information in $\rho$ as measured by quantum information entropies.  To show that these volumes behave like information entropies, we compare them to the von Neumann and linear entropies for states whose symmetries are given by $SO(3)$, $SU(2)$, and $SO(N)$. We also show that our procedure reproduces features of Boltzmann's original coarse-graining by showing that the majority of phase space consists of states near or at equilibrium. As a consequence of this coarse-graining, it is shown that an inherent flag variety structure underlies composite Hilbert spaces. 
\end{abstract}

\maketitle

\thispagestyle{fancy}

\section{Introduction}
\label{sec:intro}
\noindent The fundamental concept that lays the theoretical foundation of both statistical mechanics and information theory is entropy.  In statistical mechanics, the thermal/Boltzmann entropy is defined as
\begin{equation}
\label{eq:boltz}
S_B=k \log W,
\end{equation}
where $k$ is Boltzmann's constant and $W$ is the number of micro-states a system could be in that are consistent with macro-data.  This represents an \black{observer's} ignorance about the actual state of the system.  The larger $W$ is, the more possible states there are, the greater chance an observer is incorrect when randomly guessing which micro-state is responsible for the known macro-data.  Alternatively, information/Shannon entropy is defined as
\begin{equation}
\label{eq:shannon}
S=-\sum_{i=1}^n{p_i \log{p_i}},
\end{equation}
where $\vec{P}=\{p_{i}\}$ is the probability distribution associated with the $n$ possible outcomes $\{x_{i}\}$ of a random variable $X$.  This represents an \black{observer's} ignorance about a measurement outcome.  For example,  if $\vec{P}=\{1,0_2,0_3,...,0_n\}$, then $S=0$ and one is certain $x_1$ will be observed.  Conversely, if $\vec{P}=\{1/n,1/n,...,1/n\}$, then $S=S^{max}\black{=\log n}$ and one is maximally ignorant about what will be observed.  Although the relationship between Eqs.~\ref{eq:boltz} and~\ref{eq:shannon} is not \black{readily} obvious, the underlying concept that ties them together is ignorance.  Note that here, and throughout the paper all logarithms are taken to be base two to measure entropy in bits.  
  
The first connection between thermal and information entropies was given in 1929 by Szilard's single molecule heat engine~\cite{szilard},
which was designed to analyze Maxwell's demon~\cite{maxdemon}. Szilard showed that the decrease in entropy of the reservoir due to the extraction of energy by the molecule from doing work can be accounted for by the increase in entropy of the demon when it observes if the molecule is on the left or right side of the volume.  By computing the work done by the molecule, Szilard concluded that the change in entropy due to observation must be at least $k \log{2}$.  This shows that the minimum change in entropy of the demon is equal to one bit of information, thus connecting information and thermal entropies.  The relationship between thermal and information entropy was further established by Brillouin~\cite{brillouin} using negentropy~\cite{whatlife,brillbook}.

Brillouin's contribution to the analysis of Maxwell's demon was to include a formal analysis of observation.  For the demon to operate the door, it must observe which molecules are moving fast and which are moving slow.  Therefore, he modified the system to include radiation from a light bulb which scatters off the gas molecules that can be observed by the demon.  As a result of absorbing a scattered photon, the demon gains information about the system that allows it to reduce the number of possible micro-states $W_0$ in which the system could be.  The reduction from $W^0$ to $W^1=W^0-w$, where $w<<W^0$ is related to the information gained by the demon, reduces the thermal entropy from $S^{0}_B=k \log{W^0}$ to $S^1_B=k \log{W^1}$.  Assuming the demon had no knowledge about the state of the system prior to measurement, $S^{0}_B=S^{max}_B$ and the negentropy is defined as
\begin{equation}
\label{eq:negentropy}
I=S^{max}_B - S^1_B.
\end{equation}
This implies that negentropy quantifies information.  This is in contrast to entropy which quantifies ignorance due to a lack of information.  By relating the reduction of possible micro-states to the gain in information from observation, Brillouin directly related classical information to classical thermal entropy.  

Brillouin's approach to connecting information and thermal entropy using negentropy is widely accepted~\cite{physmax}.  In this paper, we follow the tradition of Brillouin and connect quantum information entropies to quantum thermal entropy.  To do this, we introduce \black{a canonical} quantum coarse-graining that is analogous to Boltzmann's original coarse-graining.  We do not define a quantum Boltzmann entropy.  Instead, we focus on the relationship between quantum information entropy and the multiplicity of micro-states from coarse-graining. 

\subsection{Coarse-Graining}
\noindent Coarse-graining is a data compression process that constructs lower (macro) resolution models of a system from higher (micro) resolution models by coalescing higher resolution data into lower resolution data.  This loss of data results in ignorance that is quantified by some notion of entropy.  Since the process of compressing data is not unique, there are a variety of coarse-graining descriptions both classically~\cite{castiglione08} and quantum mechanically~\cite{busch93,gellmann07}.  In quantum mechanics, one must contend with nonclassical peculiarities such as superposition~\cite{kabernik18} and the lack of commutation of conjugate degrees of freedom~\cite{safranek19A} which make counting arguments to define phase space volumes difficult.  These volumes quantify ignorance that results from coarse-graining~\cite{lloyd06}.

\begin{figure}[h]
\centering
\includegraphics[width=\columnwidth]{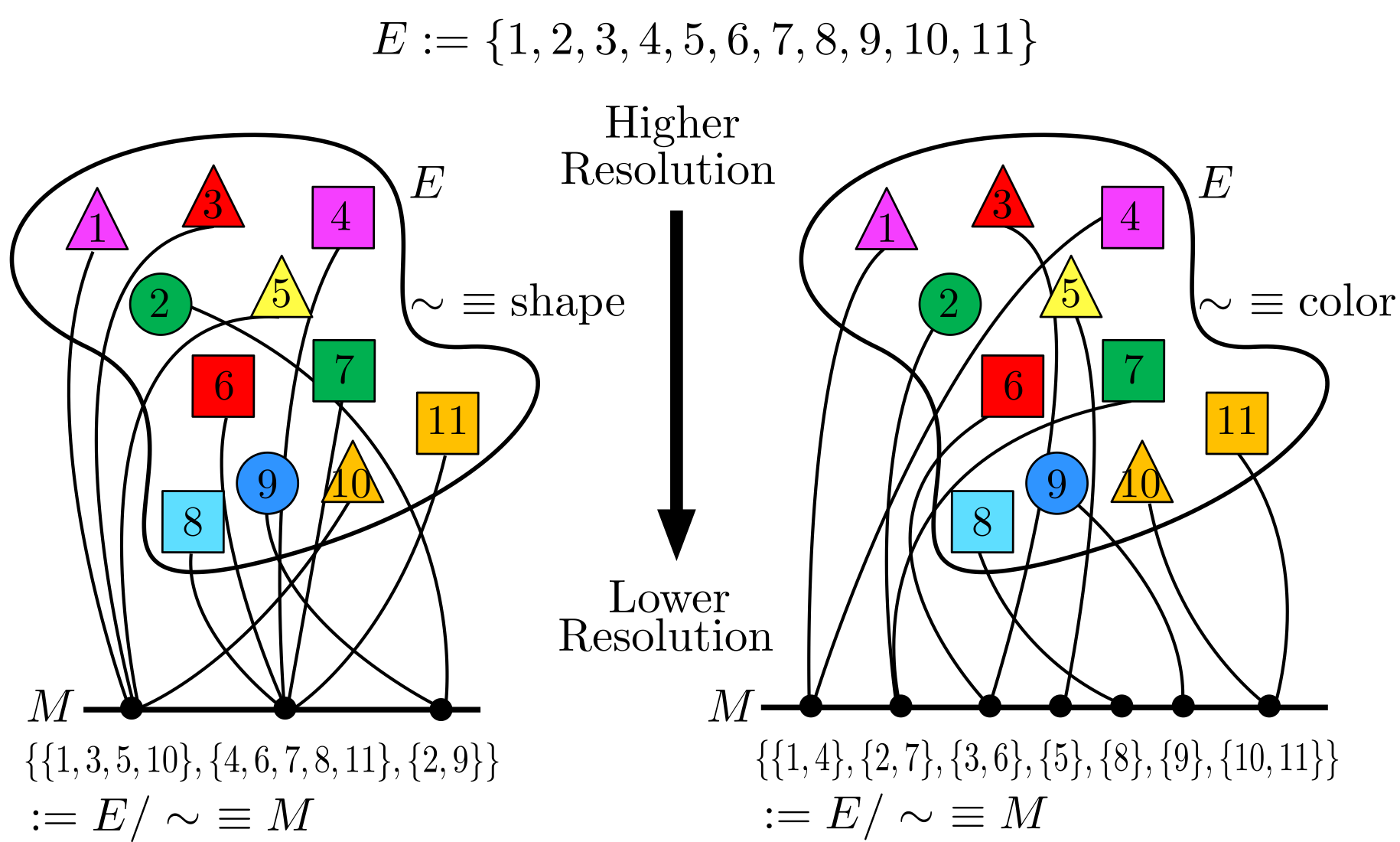}
\caption{
\black{Illustration of}
coarse-graining in terms of equivalence relations and equivalent classes.  This figure is also consistent with a fiber bundle perspective~\cite{geometryq}. The  bundle/total space in which the fibers are defined is denoted as $E$, while the quotient space, where the equivalence classes lie, is given by $M$.  Notice how different equivalent relations give different equivalent classes and thus different coarse-grainings.  In the left figure, the equivalence relation, $\sim$, is shape, while on the right it is color.  In both cases, we move from a higher resolution space that contains information about shape and color, to a lower resolution space where information about color or shape is lost.  For a point $m \in M$, there exists a set $F_m \subset E$ whose elements all map to $m$ under projection $\pi:E \rightarrow M$; $F_m$ is the fiber of $m$.  The elements of $F_m$ are the micro-states of macro-state $m$.  Since there is a one-to-one correspondence between $F_m$ and  $m$, macro-states can also be understood as $F_m$ from the perspective of $E$.}
\label{fig:maincoarse}
\end{figure}

For a coarse-graining to be well defined, there must be a clear notion of equivalence between micro-states.  That is, they are uniquely determined by equivalence relations that divide higher resolution data into disjoint sets called equivalent classes.  The fundamental nature of coarse-graining can be understood in terms of fiber bundles as seen in Fig.~\ref{fig:maincoarse}.  
Each element of an equivalence class is a micro-state, and each equivalence class is a macro-state.  Essentially, the equivalence relations function as constraint data (shared information) that all micro-states of a macro-state must share.  The shared information is the symmetry between micro-states.  As in Fig.~\ref{fig:maincoarse}, different equivalence relations result in different coarse-grainings.  This lack of uniqueness is central to the problem of defining a clear notion of quantum coarse-graining.  

In quantum, there have been many approaches to coarse-graining.  The prescription for each procedure depends on the motivation.  For example, in \cite{dimatteo17}, the authors coarse-grain an $N$-qubit system with the motivation of simplify the tomography process.  Their procedure is defined w.r.t a coarse-grained quasiprobability Wigner function that emerges from the coarse-grained phase space. The symmetries are specified by a partitioning of the so-called line structure of the phase space where thin lines in the fine phase space are grouped together to form thick lines in the coarse phase space.  In \cite{duarte17}, the authors are motivated by studying simplified dynamics of quantum systems using coarse-graining.  Their coarse-graining is defined using dimension-reducing completely positive and trace preserving (CPTP) linear maps.  In \cite{singh18}, the authors are also motivated by reducing the representation of quantum states to lower dimensional Hilbert space.  The goal is to capture the most relevant data within the system.  To do this, they have a two step procedure.  First, they employ principle component analysis (PCA) to identify a nonredundant basis and compress the dimensionality of the Hilbert space. Second, they truncate the last few PCA terms whose relevance in reconstructing the quantum state is negligible.   In \cite{kabernik18}, the author is again motivated by reducing the complexity of dynamics through coarse-graining.  To do this, they implement a generic state-space coarse-graining in addition to a secondary coarse-graining based on symmetry groups compatible with the generators of dynamics.  For a presentation on the notion of coarse-graining by symmetry, we refer to~\cite{faist16}. Furthermore, for a quantitative description of a quantum system subject to a particular symmetry, we refer to the work by Bartlett and collaborators on the so-called reference frames formalism \cite{bartlett07}.

As mentioned before, entropy quantifies ignorance. In the framework of quantum coarse-graining, a notion of coarse-grained entropy was originally proposed by Safranek and collaborators in~\cite{safranek19A}. 
\black{In this latter paper 
Safranek and collaborators proposed the notion of observational entropy
building upon the coarse-grained projection formalism by von Neumann \cite{jvoncoarse}, along with the introduction of coarse-grained entropies as presented in~\cite{wehrl78, brun99}.} 
This is a form of coarse-grained entropy that can be regarded as a generalization of the Boltzmann entropy to quantum mechanics \cite{qcoarse}. The entropy was termed \textquotedblleft observational\textquotedblright\ since the coarse-graining procedure can be described in terms of a sequence of measurements that can be selected in a free manner by an observer. For a detailed discussion on classical and quantum observational entropy, along with its links to the Boltzmann entropy, the Gibbs entropy, the thermodynamic entropy, and the second law of thermodynamics, we refer to~\cite{qcoarse,safranek20}.

The coarse-graining defined here is motivated by constructing a quantum analogue of Boltzmann's original coarse-graining. Unlike~\cite{qcoarse,safranek20}, our coarse-graining is independent of any notion of observation.  Instead, the symmetries are Lie group symmetries inherent to the Hilbert spaces in which states are defined. The resulting coarse-graining structure always exists regardless of observer or dynamics and it is unique. This is in contrast to procedures such as the von Neumann projector approach since all demarcations of Hilbert space into disjoint subspaces are valid and thus not unique.  As far as we know, no other proposed quantum coarse-graining procedure gives a unique coarse-graining.  

In our coarse-graining, micro-states are defined as elements of the set of purifications $F_{\rho}\equiv\{ |\bar{\Gamma}^{\rho}(\vec{\xi})\rangle\}$, which is continuously parameterized by $\vec{\xi}$, for a macro-state $\rho$; the set of purifications $F_{\rho}$ is the fiber of $\rho$.  The fibers are generated by transitive Lie group actions on $\rho$ which guarantees that each set of purifications is uniquely determined by \black{the eigenvalues of} $\rho$.  The Lie groups that generate $F_{\rho}$ are the symmetries of the space in which $\rho$ is defined.  In using the Lie group symmetries (uniquely defined by $\rho$) as equivalence relations between micro-states of a particular macro-state, we can define a unique coarse-graining we call canonical.  

Our canonical quantum coarse-graining has several unique properties.  (1) The phase spaces in which our micro-states are defined are the composite Hilbert spaces $\mathcal{H}_R \otimes \mathcal{H}_A$ where $\rho \in \mathcal{H}_A$ and $\mathcal{H}_R$ is a reservoir space.  This gives a natural setting for modeling thermalization within a closed composite system consisting of an environment and state $\rho$, which we will discuss in Sec.~\ref{sec:physexamp}.  We use the term phase space in the broadest sense to mean the space in which all possible states of a system are represented.  This is consistent since purifications are equivalent up to the macro-data in $\rho$.   (2) It allows us to make a clear relationship between the information entropy of $\rho$ and the ignorance associated with its fiber $F_{\rho}$.  We show that pure states have zero volume while maximally mixed states have maximal volume, and that the volume is monotonic w.r.t information entropy.  This gives a coarse-graining of phase space that is consistent with Boltzmann's original coarse-graining where macro-states with greater information entropy take up a larger fraction of volume in phase space.  This property was essential to Boltzmann's arguments when defending his H-theorem~\cite{goldstein,stanford}. (3) It gives a natural partitioning of phase space into disjoint sets since, by definition, all elements of $F_{\rho}$ map to a single $\rho$ in $\mathcal{H}_A$.  (4) For each $\rho$, one can access the complete set of purifications using the set of unitaries associated with the symmetries of $\mathcal{H}_R$.  That is, our symmetries are defined with respect to Lie group actions.  This allows the construction of differential manifolds associated with each set of purifications whose volume is the phase space volume associated with macro-state $\rho$.  (5) The fibers generated by Lie group symmetries are related to gauge where the set of purifications represents the gauge freedom when choosing how to model the system.  (6) A flag variety structure emerges naturally from this coarse-graining which connects our analysis to fields that use flags such as quantum field theory~\cite{flagphysics}.  

The paper is structured as follows.  In Sec.~\ref{sec:back}, we give some background on the Boltzmann entropy.  This will put our coarse-graining in context with the original coarse-graining proposed by Boltzmann.   In Sec.~\ref{sec:cqcg}, we will define our micro- and macro-states, the metric components, and volume of the surfaces of ignorance.  In Sec.~\ref{sec:interp}, we give an interpretation of the volume in terms of information entropy and negentropy.  In Sec.~\ref{sec:volexamp} we present examples of our volume using states described by $SO(3)$, $SU(2)$, and $SO(N)$ symmetries.  This demonstrates how the volume of the surfaces of ignorance behaves like an information entropy, and that our coarse-graining reproduces features of Boltzmann's original coarse-graining.  In Sec.~\ref{sec:flags}, we provide qualitative discussions on the inherent flag variety structure~\cite{hilbertflag} that arises from this coarse-graining and physical examples.  The discussion on flags is meant to demonstrate how they ground the canonical nature of our coarse-graining by providing relationships between phase spaces of subsequently greater dimensions as one purifies density operators ad infinitum.  We conclude the paper in Sec.~\ref{sec:conclusion}. 
  
\section{Boltzmann's Coarse-Graining}
\label{sec:back}
\noindent Although it is not the \black{primary} purpose of this paper to define a quantum Boltzmann entropy, it does lay the foundation \black{for doing so}.  For a clear quantum Boltzmann entropy to be defined, one must have a \black{well-defined} coarse-graining procedure and a method for quantifying the resulting ignorance.  Traditionally, this ignorance is quantified by taking the logarithm of the volume whose elements are treated equivalently by the symmetries that characterize the coarse-graining.  The goal of this paper is only to define a coarse-graining and examine the resulting phase space volumes.  By examining these volumes, we connect quantum information and thermal entropies though a precise definition of the quantum Boltzmann entropy is left for future research.  It is important however to discuss our analysis in the context of the original coarse-graining proposed by Boltzmann.  Namely we want to emphasize the similarities regarding the size of volumes in phase space and equilibrium states.

One of the significant consequences of Boltzmann's original entropy was the observation that micro-states with less disorder are less likely to be observed than those with greater disorder.  This is due to the combinatoric symmetries  of possible distributions on the single particle phase space also known as $\mu$-space. 

\begin{figure}[h]
\centering
\includegraphics[width=\columnwidth]{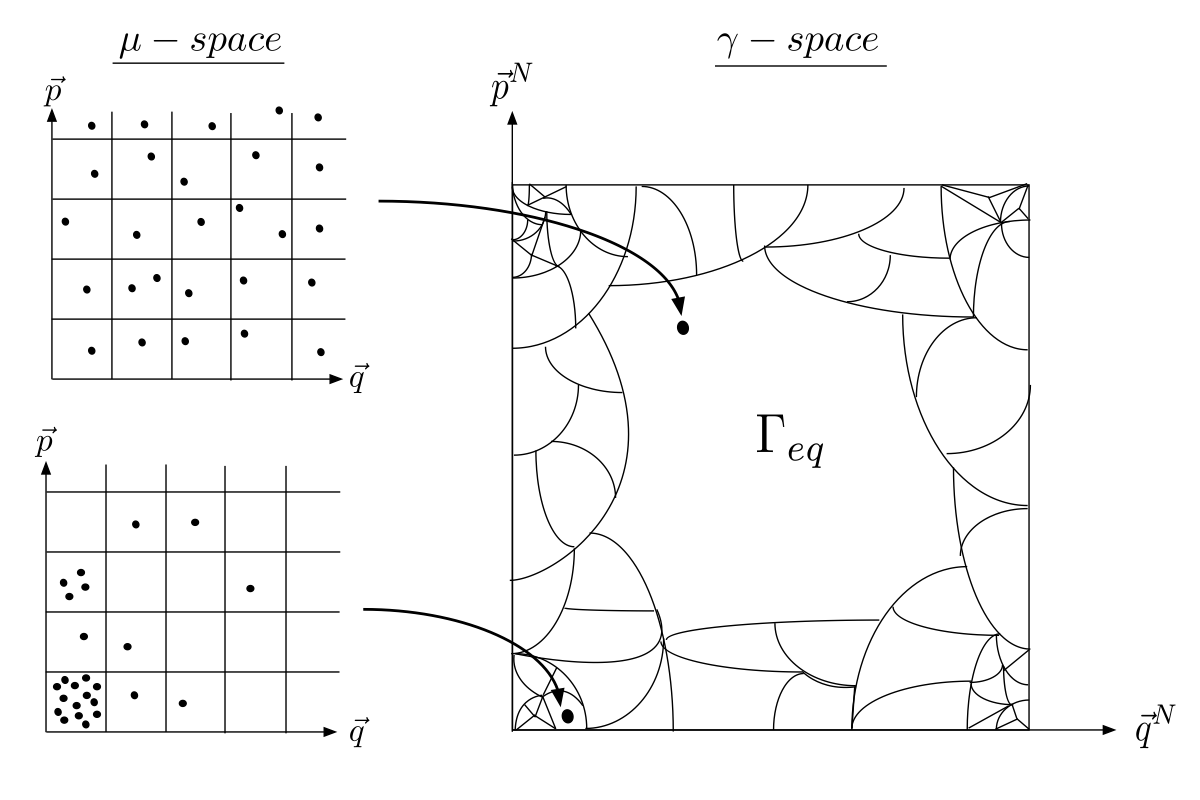}
\caption{
\black{Illustration of}
the relationship between $\mu$-space and $\gamma$-space.  It also shows Boltzmann's original approach to coarse-graining.  By dividing $\mu$-space into equal cells, Boltzmann simply counted the number of particles in each cell.  With this division of $\mu$-space, $\gamma$-space is demarcated into mutually exclusive sets as shown in the right figure.  The size of each set is determined by the number of combinatorially equivalent sets $Z_j$.  The set $\Gamma_{eq}$ contains all sets $Z_j$ that are consistent with a gas in equilibrium.  These states make up the majority of the $\gamma$-space volume; this fact was essential to Boltzmann's original analysis. States with low entropy, like the distribution depicted by the bottom left figure, have less equivalent combinations and thus represent a smaller volume in $\gamma$-space.}
\label{fig:mugamma}
\end{figure}

The original system for which Boltzmann's entropy was defined was an isolated gas in volume V with N particles and total fixed energy E.  To define his entropy, Boltzmann divided $\mu$-space into equal position and momentum bins as seen in the left imagine of Fig.~\ref{fig:mugamma}.  Assuming there are m cells, one can define occupation sets 
\begin{equation}
\label{eq:Z}
Z_j:=\{n_1,n_2,...,n_m\}_j
\end{equation} 
where $n_k$ is the number of particles that occupy the kth cell, $p_k=n_k/N$ is the probability of finding a particle in the kth cell, and $\sum^m_{i=1}{n_i}=N$.  Given the set of all occupation sets, $\{Z_j\}_{j\in J}$, where $J$ is an index set, there exist subsets $J_a \subset J$ where each $j \in J_a$ index combinatorially equivalent $Z_j$'s.  For example, let $J_1$ contain all indices that index $Z_j$'s with all particles found in a single cell, and let $Z_1:=\{N,0_2,...,0_m\}$ and $Z_2:=\{0_1,N,...,0_m\}$; then $j=1,2 \in J_1$.  The combinatorial symmetries captured by $J_a$ are the symmetries that define Boltzmann's coarse-graining. 

The combinatorial symmetries are the basis for coarse-graining the full phase space, which is also known as $\gamma$-space.  In this case, $j \in J_a$ parameterizes micro-states belonging to macro-states parameterized by $a$.  This is depicted by the right image of Fig.~\ref{fig:mugamma}.  As Boltzmann showed, macro-states consisting of micro-states with less disorder make up significantly less volume than those near or at equilibrium.  This is an important feature of Boltzmann's work since it was the basis of defending his H-theorem~\cite{goldstein,stanford}.  It was also emphasized as significant in~\cite{goldplus} when attempting to define a quantum Boltzmann entropy.  As we will show, the partitioning of the phase space $\mathcal{H}_{RA}$ into disjoint sets of macro-states by our canonical quantum coarse-graining is consistent with this feature of Boltzmann's original analysis.

\section{Canonical Quantum Coarse-Graining}
\label{sec:cqcg}
\noindent In this section, we define our coarse-graining and the resulting ignorance.  We begin by defining the micro- and macro-states.  Subsequently, we give the explicit formulation of these states.  Once the states are defined, we construct metric components of the differential manifold constructed from the continuous parameterization of the micro-states.  Finally, we quantify the ignorance by giving an expression for the volume of this manifold.  


\subsection{Micro- and Macro-States}
\noindent To define our coarse-graining, we begin by defining our micro- and macro-states. The higher resolution data is defined in the composite Hilbert space $\mathcal{H}_{RA}\equiv \mathcal{H}_R \otimes \mathcal{H}_A$ (which is analogous to $\gamma$-space), and the lower resolution data is defined in the base space $\mathcal{H}_A$ (which is analogous to $\mu$-space). Here, the equivalence relations are defined by the group symmetries $G$ inherent to $\mathcal{H}_R$ such that $\mathcal{H}_A = \mathcal{H}_{RA}/G$.  For each density operator $\rho \in \mathcal{H}_A$, there exists a fiber $F_{\rho}\equiv\{|\bar{\Gamma}^{\rho}(\vec{\xi})\rangle\} \subset \mathcal{H}_{RA}$ such that
\begin{equation}
\label{eq:microset}
\rho=\Tr_R \left[|\bar{\Gamma}^{\rho}(\vec{\xi})\rangle \langle \bar{\Gamma}^{\rho}(\vec{\xi})|\right],
\end{equation}
where $\vec{\xi}$ is a continuous parameterization of $F_{\rho}$.  The micro-states are the elements of the fiber $F_{\rho}$, and the constraint information that characterizes the macro-state $\rho$ are the eigenvalues $\{\lambda^i_{\rho}\}$ and eigenvectors $\{|\lambda^i_{\rho}\rangle\}$.  The macro-states can be seen from two perspectives.  In $\mathcal{H}_A$, it is the density operator $\rho$. But if one is only looking at the coarse-graining from the perspective of $\mathcal{H}_{RA}$, then the macro-state is the set $F_{\rho}$ that is uniquely determined by $\rho$.  That is, the fibers are analogous to the disjoint sets of $\gamma$-space as seen in Fig.~\ref{fig:mugamma}.   

Since the symmetries that generate the fiber are Lie group symmetries, there exists a differential manifold $\mathcal{M}_{\rho} \subset \mathcal{H}_{RA}$ associated with it \black{constructed from $F_\rho$ and $\rho$}.  These manifolds are the surfaces of ignorance and their volumes quantify the ignorance that emerges from coarse-graining.  With our micro- and macro-states defined, we now follow the prescription given in section 5 of Wilde's ``Quantum Information Theory"~\cite{wilde} to generate $F_{\rho}$.

\begin{figure}[h]
\centering
\includegraphics[width=\columnwidth]{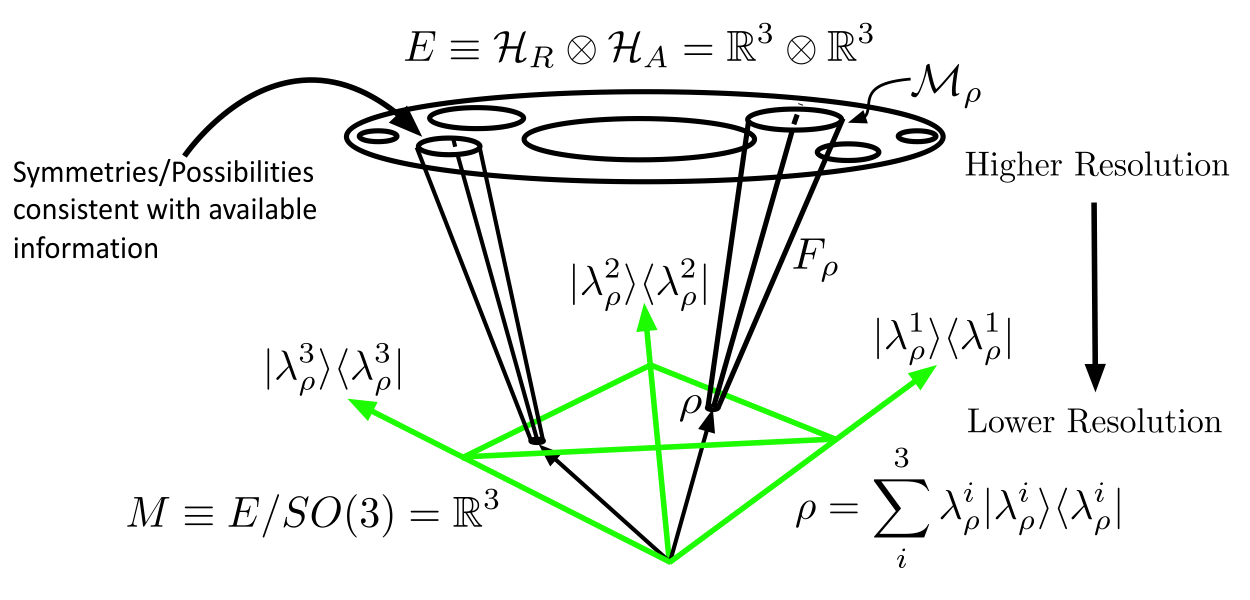}
\caption{
\black{Depiction} of micro- and macro-states for 3-level systems with complex phase angles set to zero.  There is nothing special about this choice of states, but it does make visualization easier.  As in Fig.~\ref{fig:maincoarse}, the micro-states are represented as fibers in the higher resolution space $E$, and the macro-states are represented as points in the lower resolution space $M$.  The equivalence relations are defined by the symmetries of $\mathcal{H}_R$ and $\mathcal{H}_A$, which are $SO(3)$.  Also note that the size of the fibers is related to the location of $\rho$ on the probability simplex.  As a state becomes less pure, the volume of its fiber increases.  As will be shown later, the volume of $\mathcal{M}_{\rho}$ is related to ignorance as measured by the information entropy of $\rho$.
}
\label{fig:fib2}
\end{figure}

To construct $F_p$, we begin with \black{a canonical} purification
\begin{equation}
\label{eq:canonical}
|\phi^{\rho}\rangle_{RA} = (\hat{1}_R \otimes \sqrt{\rho_A})|\Gamma \rangle_{RA}
\end{equation}
where $\hat{1}_R$ is the identity operator in $\mathcal{H}_R$, 
\begin{equation}
\label{eq:gamma}
|\Gamma \rangle_{RA} = \sum^d_{i=1}{|i\rangle_R |i\rangle_A} 
\end{equation}
is the unnormalized Bell state, and $d$ is the dimension of the system.  From here, one can access all purifications by applying unitary transformations associated with the symmetries of $\mathcal{H}_R$ to Eq.~\ref{eq:canonical}.  This gives,
\begin{equation}
\label{eq:gammabar}
|\bar{\Gamma}^{\rho}(\vec{\xi})\rangle_{RA} = (U_R(\vec{\xi}) \otimes \hat{1}_A)|\phi^{\rho} \rangle_{RA}=(U_R(\vec{\xi})\otimes \sqrt{\rho_A})|\Gamma \rangle_{RA}.
\end{equation}
In Fig.~\ref{fig:fib2}, we depict the differential manifold $\mathcal{M}_{\rho}$ 
which is derived from $F_{\rho}$ and $\rho$.  This figure is analogous to Fig.~\ref{fig:maincoarse} where $E \equiv \mathcal{H}_{RA}$, and $M \equiv \mathcal{H}_A$.  Next, we compute the metric and volume of $\mathcal{M}_{\rho}$.

\subsection{Surfaces of Ignorance: Metric Components and Volume}
\noindent To compute the metric components and volume associated with \black{$\mathcal{M}_{\rho}$}, we construct its first fundamental form using a Taylor expansion of Eq.~\ref{eq:gammabar}.  Expanding around parameters $\vec{\xi}$ using $\vec{\xi}_0$, the displacement vector is given by $d\vec{\xi}=\vec{\xi}-\vec{\xi}_0$.  Taking the Taylor expansion of $|\bar{\Gamma}(\vec{\xi})\rangle$ to first order, and bringing the zeroth order term to the l.h.s, the differential is given by
\begin{equation}
\label{eq:differential}
|d\bar{\Gamma}\rangle \equiv |\bar{\Gamma}(\vec{\xi}_0+d\vec{\xi})\rangle-|\bar{\Gamma}(\vec{\xi}_0)\rangle=\sum^n_{i=1} |\bar{\Gamma}_{,\xi_i}\rangle d\xi_i
\end{equation}
where $n$ is the number of parameters of the symmetry.  For simplicity of notation, we dropped the superscript $\rho$ from $|\bar{\Gamma}^{\rho}(\vec{\xi})\rangle$.  Since we are working in $\mathcal{H}_{RA}$, and all of our states are pure, the scalar product is well defined.  The metric components induced by the scalar product are given by the first fundamental form
\begin{equation}
\label{eq:firstfund}
ds^2=\langle d\bar{\Gamma}|d\bar{\Gamma}\rangle = \langle \bar{\Gamma}_{,i}|\bar{\Gamma}_{,j}\rangle d\xi_i d\xi_j
\end{equation}
where the metric components are $g_{ij}=\langle \bar{\Gamma}_{,i}|\bar{\Gamma}_{,j}\rangle$.  From Eq.~\ref{eq:firstfund}, the volume element is $dV=\sqrt{\hbox{Det}[\bold{g}]}\ d\xi_1 d\xi_2...d\xi_n$ and the volume is
\begin{equation}
\label{eq:volume}
\black{V\equiv V_{\mathcal{M}_{\rho}} =} \int_{\xi_1}{\int_{\xi_2}{...\int_{\xi_n}{dV}}}.
\end{equation}
Next we discuss the relationship between missing information and possible micro-states in the context of negentropy.

\section{Physical Interpretation}
\label{sec:interp}
\noindent In quantum information theory, information entropies measure a \black{state's} deviation from purity.  The more pure a state is, the more information it has and the more certain an observer is that the statistics of the system that is being measured are consistent with a specific pure state.  The macro-data, $\{\lambda^i_{\rho}\}$ and $\{|\lambda^i_{\rho}\rangle\}$, in $\rho$ constrains the set of possible pure states responsible for observed statistics.  The less pure a state is, the less information there is to constrain the set of possible pure states.  This is the essence of Brillouin's connection between thermal and information entropies using negentopy.  

For our coarse-graining, the lack of information/constraints manifests itself as a greater volume of $\mathcal{M}_{\rho}$ where each point is a pure state of greater dimension that completes the missing information in $\rho$.  If one discovered which purification was responsible for $\rho$, then according to the negetropy, they would go from having initial information $I^0 = \log{d_A} - S(\rho)$ to $I^1 = \log{d_{RA}}$ where $d_A$ and $d_{RA}$ are the respective dimensions of $H_A$ and $H_{RA}$ and $S(\rho) \in [0,\log{d}]$ is a quantum information entropy.  This implies an information gain of
\begin{equation}
\label{eq:deltainfo}
\Delta I = I^1-I^0=\log{d_{RA}}-\left(\log{d_A} - S(\rho) \right).
\end{equation}
This gain in information is not possible without additional observations.  In the absence of information identifying which purification in the fiber is the true microstate, a uniform (maximally mixed) distribution of all possible purifications should be assumed.  This is the principle of insufficient reason~\cite{insufficient} and the meaning of treating the set of possible micro-states symmetrically.  To place anything other than a maximally mixed distribution on the set of purifications would, as stated by Jaynes~\cite{jaynes}, ``amount to arbitrary assumption of information which by hypothesis we do not have".  

As the volume of $\mathcal{M}_{\rho}$ increases, one is less certain about which purification completes the missing information and is thus more ignorant about the true state of $\rho$ as a subsystem of a composite system.  But, one has the freedom to choose any purification to model the statistics of the subsystem since they are all consistent with $\rho$.  This freedom to choose a purification is \black{a gauge} freedom where the symmetry group associated with the gauge is the Lie group that captures the symmetries of $\mathcal{H}_R$.  Since $\mathcal{H}_{RA}$ is the phase space of $\mathcal{H}_A$, parameters $\vec{\xi}$ are the phase angles, and fixing them is equivalent to fixing the gauge.  

As we will show in the next section, the volume given by Eq.~\ref{eq:volume} has some necessary properties of a quantum information entropy~\cite{genentropy}.  We show that it is an upper bound of the linear and von Neumann entropies. It is zero for pure states \black{(since there is no missing information)},
maximal on maximally mixed states, monotonic w.r.t purity, and concave (so that mixing cannot decrease entropy).  We also show for states whose symmetries are given by $SO(3)$ that the coarse-graining of $\mathcal{H}_{RA}$ into disjoint sets is consistent with features of Boltzmann's original coarse-graining.  That is, we show that the majority of the volume of phase space is subsumed by states near or at equilibrium.

\section{Volume Examples}
\label{sec:volexamp}
\noindent In this section, we will present explicit examples of the volume associated with density operators $\rho \in \mathcal{H}_A$.  Our main focus will be on 3-level systems using diagonalized density operators with complex phase angles set to zero.  This example is a good introduction to our canonical coarse-graining since $\mathcal{H}_R =\mathcal{H}_A = \mathbb{R}^3$.  This implies that the symmetry group that produces $\{|\bar{\Gamma}^{\rho}(\vec{\xi})\rangle\}$ is $SO(3)$ and its representation is simply the Euler transformations in $\mathbb{R}^3$.  We also present the results for $SU(2)$ and a generalized form of the volume for $SO(N)$.  
\black{Before} we give our examples, we first define arbitrary unitary transformations for $d$-level systems.

Following the prescription in~\cite{randomu}, any arbitrary $d$-dimensional unitary transformation can be written as successive transformations of 2-dimensional subspaces.  Let $E^{(i,j)}(\phi_{ij},\psi_{ij},\chi_{ij})$ be an arbitrary transformation about the $(i,j)$-plane.  Its components are 
\begin{eqnarray}
E^{(i,j)}_{kk}&=&1\ \ \ \ \ \ k=1,...,d\ \ \ \ \ \ k\neq i,j \nonumber \\
E^{(i,j)}_{ii}&=&e^{i \psi_{ij}} \cos{\phi_{ij}}\nonumber \\
E^{(i,j)}_{ij}&=&e^{i \chi_{ij}} \sin{\phi_{ij}} \\
E^{(i,j)}_{ji}&=&-e^{-i \chi_{ij}} \sin{\phi_{ij}}\nonumber \\
E^{(i,j)}_{jj}&=&e^{-i \psi_{ij}} \cos{\phi_{ij}}\nonumber
\end{eqnarray}
and zero everywhere else.  The superscript indices $(i,j)$ index the 2-D plane about which the transformation is applied, and the subscripts are the nonzero matrix indices. From here, one can construct successive transformations
\begin{eqnarray}
&E_1& = E^{(1,2)}(\phi_{12},\psi_{12},\chi_{12})\nonumber \\
&E_2& = E^{(2,3)}(\phi_{23},\psi_{23},0)E^{(1,3)}(\phi_{13},\psi_{13},\chi_{13})\nonumber \\
&&. \\
&&.\nonumber \\
&&.\nonumber \\
&E_{N-1}& = E^{(N-1,N)}(\phi_{N-1,N},\psi_{N-1,N},0)\nonumber \\ 
&\ &\ \ \ E^{(N-2,N)}(\phi_{N-2,N},\psi_{N-2,N},0)\nonumber \\
&\ &\ ...\ E^{(1,N)}(\phi_{1N},\psi_{1N},\chi_{1N})
\end{eqnarray}
and finally an arbitrary $U(N)$ transformation
\begin{equation}
\label{eq:UN}
U = e^{i\alpha}E_1E_2\ ...\ E_{N-1}.
\end{equation}
With the arbitrary unitaries defined, we now present our examples.

\subsection{Example: $SO(3)$}
\subsubsection{Computing Volume}
\noindent Given Eq.~\ref{eq:UN}, the unitaries associated with $SO(3)$ are given by choosing $d=3$ and $\alpha=\psi_{ij}=\chi_{ij}=0$ for all $i$ and $j$.  This removes all complex phases, which leaves parameters $\vec{\xi}=\{\phi_{12},\phi_{13},\phi_{23}\}$ where $\phi_{12},\phi_{13}\in [0,2\pi]$ and $\phi_{23} \in [0,\pi]$.  The resulting unitaries are given by
\begin{widetext}
\begin{equation}
U(\phi_{12},\phi_{13},\phi_{23})=
\begin{bmatrix}
\cos \phi_{12}\cos{\phi_{13}} -\sin{\phi_{12}} \sin{\phi_{13}} \sin{\phi_{23}} & \cos{\phi_{23}}\sin{\phi_{12}} & \cos{\phi_{12}}\sin{\phi_{13}}+\cos{\phi_{13}}\sin{\phi_{12}}\sin{\phi_{23}} \\
-\cos{\phi_{13}}\sin{\phi_{12}}-\cos{\phi_{12}}\sin{\phi_{13}}\sin{\phi_{23}} & \cos{\phi_{12}}\cos{\phi_{23}} & -\sin{\phi_{12}}\sin{\phi_{13}}+\cos{\phi_{12}}\cos{\phi_{13}}\sin{\phi_{23}} \\ 
-\cos{\phi_{23}}\sin{\phi_{13}} & -\sin{\phi_{23}} & \cos{\phi_{13}}\cos{\phi_{23}}  
\end{bmatrix}.
\end{equation}
\end{widetext}
Since $U(\phi_{12},\phi_{13},\phi_{23})$ generates the symmetries of both $\mathcal{H}_A$ and $\mathcal{H}_R$, we will use the sub-labels $A$ and $R$ to keep track of which space $U$ is acting upon.  

\begin{figure}[h]
\centering
\includegraphics[width=2.6in]{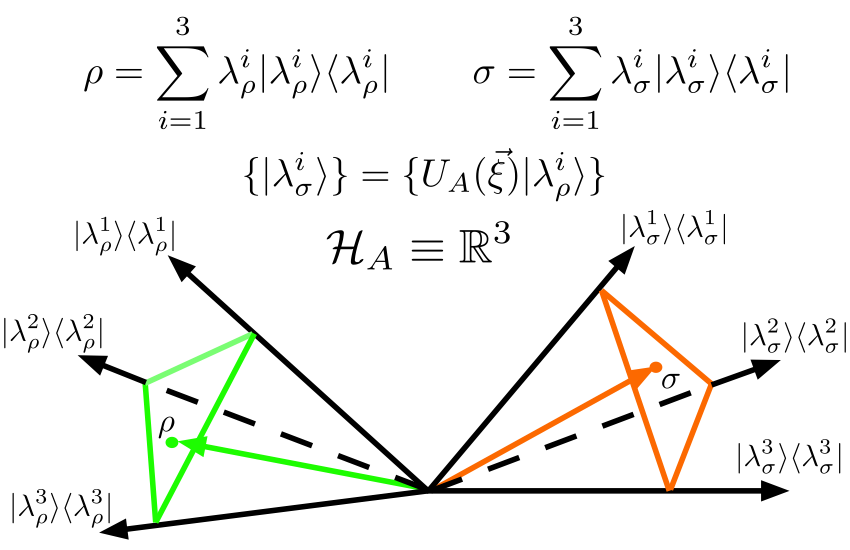}
\caption{In this representation of $\mathcal{H}_A$, pure states (and their projectors) are represented by unit vectors and maximally mixed states are vectors with magnitude $1/\sqrt{3}$.  All density operators are assumed to be diagonalized.  As such, they are defined as linear combinations of orthonormal bases where the components (eigenvalues) must sum to 1.  This can be seen for $\rho$ and $\sigma$ which are represented as points on probability simplicies.  Here we see that the eigenbases of $\sigma$ are related to the eigenbases of $\rho$ by unitary transformation.}
\label{fig:tor3}
\end{figure}
 

For completeness of our formalism, which will be relevant in Sec.~\ref{sec:flags} when we discuss its generalization using flags, we include unitary transformations of $\{|\lambda^i_{\rho} \rangle\}$.  This allows us to define density operators $\sigma$ on unitarily equivalent probability simplicies  whose bases $\{|\lambda^i_{\sigma}\rangle\} =\{ U_A(\phi^{\sigma}_{12},\phi^{\sigma}_{13},\phi^{\sigma}_{23})|\lambda^i_{\rho}\rangle \}$.  Using the unitaries of $\mathcal{H}_A$, all orthonormal bases are generated and are used to define all possible probability simplicies.  This is relevant when studying quantum fidelity which is characterized by comparing density operators that do not share eigenbases.  If the eigenbases are the same, then the fidelity analysis is merely classical.  In fact, the Uhlmann-Josza fidelity can be characterized as the maximum overlap of purifications between $\rho$ and $\sigma$~\cite{wilde}.

One of the advantages of this example is the ease of visualization.  
\black{Examining} Fig.~\ref{fig:tor3}, we see the representations of $\rho=\sum_{i=1}^3{\lambda_{\rho}^i |\lambda_{\rho}^i\rangle \langle \lambda_{\rho}^i|}$ and $\sigma=\sum_{i=1}^3{\lambda_{\sigma}^i |\lambda_{\sigma}^i\rangle \langle \lambda_{\sigma}^i|}$ as points in $\mathbb{R}^3$.  Since $\rho$ and $\sigma$ are diagonalized and the sum of their eigenvalues must equal one, they can be treated like classical random variables where their eigenvalues $\{\lambda^i_{\rho,\sigma}\}$ act as classical probabilities of observing eigenvectors $\{|\lambda_{\rho,\sigma}^i\rangle\}$.  Here,   
\begin{equation}
\{|\lambda^i_{\rho} \rangle \} =
\begin{Bmatrix}
\begin{bmatrix}
1 \\
0 \\
0 \\
\end{bmatrix}
, &
\begin{bmatrix}
0 \\
1 \\
0 \\
\end{bmatrix}
, &
\begin{bmatrix}
0 \\
0 \\
1 \\
\end{bmatrix}
\end{Bmatrix},
\end{equation}
Notice that $\vec{\xi}_A=(\phi^{\sigma}_{12},\phi^{\sigma}_{13},\phi^{\sigma}_{23})$ is fixed.  As we will see when we discuss flag varieties, we have access to the unitary freedoms in $\mathcal{H}_A$ as well as $\mathcal{H}_R$ which gives us access to the entirety of the flag.

With the bases defined, the unnormalized maximally entangled state for $\rho$ and $\sigma$ are
\begin{equation}
\label{eq:maxentrho}
|\Gamma^{\rho}\rangle =\sum^3_{i=1}{|\lambda^i_{\rho}\rangle|\lambda^i_{\rho}\rangle},
\end{equation}
and
\begin{equation}
\label{eq:maxentsig}
|\Gamma^{\sigma}\rangle =\sum^3_{i=1}{|\lambda^i_{\sigma}\rangle|\lambda^i_{\sigma}\rangle},
\end{equation}
respectively.  Therefore, from Eq.~\ref{eq:gammabar}, the set of purifications for $\rho$ and $\sigma$ are
\begin{equation}
\label{eq:pureallrho}
|\bar{\Gamma}^{\rho}\rangle = (U_R\otimes \sqrt{\rho})|\Gamma^{\rho}\rangle=\sum^3_{i=1}{\sqrt{\lambda^i_{\rho}}U_R|\lambda^i_{\rho}\rangle\otimes |\lambda^i_{\rho}\rangle},
\end{equation}
and 
\begin{equation}
\label{eq:pureallsig}
|\bar{\Gamma}^{\sigma}\rangle = (U_R\otimes \sqrt{\sigma})|\Gamma^{\sigma}\rangle=\sum^3_{i=1}{\sqrt{\lambda^i_{\sigma}}U_R|\lambda^i_{\sigma}\rangle\otimes |\lambda^i_{\sigma}\rangle}.
\end{equation}

In this paper, we only compute the metric components for the surfaces of ignorance associated with the $\rho$ simplex.  This is because we are interested in analyzing volumes, and the volumes are independent of one's choice of $(\phi_{12}^{\sigma},\phi_{13}^{\sigma},\phi^{\sigma}_{23})$.  As stated previously, the purpose of including $\sigma$ is for completeness.  The difference in geometry between $\mathcal{M}_{\rho}$ and $\mathcal{M}_{\sigma}$ of unitarily related simplicies is of interest for future research.  

Writing the metric components of $\mathcal{M}_{\rho}$ explicitly, the nonzero components are
\begin{widetext}
\begin{eqnarray}
\label{eq:psipsi}
g_{\phi_{12} \phi_{12}}&=&\langle \bar{\Gamma}^{\rho}_{,\phi_{12}}|\bar{\Gamma}^{\rho}_{,\phi_{12}}\rangle=\sin^2\phi_{23}+\frac{1}{4}\left(\lambda^1_{\rho}+\lambda^2_{\rho}+3(\lambda^1_{\rho}+\lambda^2_{\rho})\cos2\phi_{23}+2(\lambda^1_{\rho}-\lambda^2_{\rho})\cos2\phi_{13} \sin^2\phi_{23}\right) \\
\label{eq:phiphi}
g_{\phi_{13} \phi_{13}}&=&\langle \bar{\Gamma}^{\rho}_{,\phi_{13}}|\bar{\Gamma}^{\rho}_{,\phi_{13}}\rangle=\lambda^1_{\rho}+\lambda^2_{\rho} \\
\label{eq:thetatheta}
g_{\phi_{23} \phi_{23}}&=&\langle \bar{\Gamma}^{\rho}_{,\phi_{23}}|\bar{\Gamma}^{\rho}_{,\phi_{23}}\rangle=\frac{1}{2}(2-\lambda^1_{\rho}-\lambda^2_{\rho}-(\lambda^1_{\rho}-\lambda^2_{\rho})\cos2\phi_{13}) \\
\label{eq:psiphi}
g_{\phi_{12} \phi_{13}}&=&g_{\phi_{13} \phi_{12}}=\langle \bar{\Gamma}^{\rho}_{,\phi_{12}}|\bar{\Gamma}^{\rho}_{,\phi_{13}}\rangle=(\lambda^1_{\rho}+\lambda^2_{\rho})\cos \phi_{23} \\
\label{eq:psitheta}
g_{\phi_{12} \phi_{23}}&=&g_{\phi_{23} \phi_{12}}=\langle \bar{\Gamma}^{\rho}_{,\phi_{12}}|\bar{\Gamma}^{\rho}_{,\phi_{23}}\rangle=-(\lambda^1_{\rho}-\lambda^2_{\rho})\cos\phi_{13} \sin\phi_{23} \sin\phi_{13}
\end{eqnarray}
\end{widetext}
where $g_{\phi_{13}\phi_{23}}=g_{\phi_{13}\phi_{23}}=0$. Taking $\sqrt{\hbox{Det}(\bold{g})}$ and integrating over the Euler angles gives
\begin{eqnarray}
\label{eq:vola}
V_{SO(3)}&=&\sqrt{(\lambda_{\rho}^1+\lambda_{\rho}^2)(\lambda^1_{\rho}+\lambda^3_{\rho})(\lambda^2_{\rho}+\lambda^3_{\rho})} \\ 
\label{eq:volb}
&=& \sqrt{(1-\lambda^1_{\rho})(1-\lambda^2_{\rho})(\lambda^1_{\rho}+\lambda^2_{\rho})}
\end{eqnarray}
where the second equality is due to the fact the eigenvalues sum to one. From Eqs.~\ref{eq:vola} and~\ref{eq:volb}, it is easy to see that $V_{SO(3)}$ is a concave function w.r.t $\vec{\lambda}_{\rho}$.

\subsubsection{Analyzing Volume}
\label{sec:volana}
\noindent In this section, we compare $V_{SO(3)}$  to the linear entropy $S_L=1-\Tr[\rho^2]$ and the von Neumann entropy $S_{VN}=-\sum^d_{i=1}{\lambda^i_{\rho}\log \lambda^i_{\rho}}$.  We also analyze the coarse-graining of $\mathcal{H}_{RA}$ and show that the distribution of phase space volumes is weighted toward states near or at equilibrium.  This result is especially interesting since we are only analyzing a $3$-dimensional system.  

Unlike Boltzmann, our coarse-graining is defined on continua of density operators and volumes. Therefore, we must discretize both to compare our results to Boltzmann.  Additionally, the analysis is completed with all three measures normalized on the interval $L=[0,1]$.  For this reason, they are denoted as $V^{\textrm{norm}}_{SO(3)}$, $S^{\textrm{norm}}_L$, and $S^{\textrm{norm}}_{VN}$ .

Examining Fig.~\ref{fig:fib2}, we see that each state $\rho$ on the probability simplex, which we will denote as $\mathcal{S}$,  has a corresponding fiber/surface of ignorance $\mathcal{M}_{\rho}$ in $\mathcal{H}_{RA}$.  Each micro-state on $\mathcal{M}_{\rho}$ represents a possible configuration of the system.  As such, $\mathcal{M}_{\rho}$ is the phase space of $\rho$, and $\mathcal{M} \equiv \bigcup_{\rho}{\mathcal{M}_{\rho}}$ is the total phase of $\mathcal{S}$.  Comparing Fig.~\ref{fig:fib2} to Fig.~\ref{fig:mugamma}, we see that $\mathcal{S}$ is analogous to $\mu$-space and $\mathcal{M}$ is analogous to $\gamma$-space.  This means that the macro-states of $\gamma$-space are analogous to the fiber/surfaces of ignorance $\mathcal{M}_{\rho}$.   
\begin{figure}[h]
\centering
\includegraphics[width=\columnwidth]{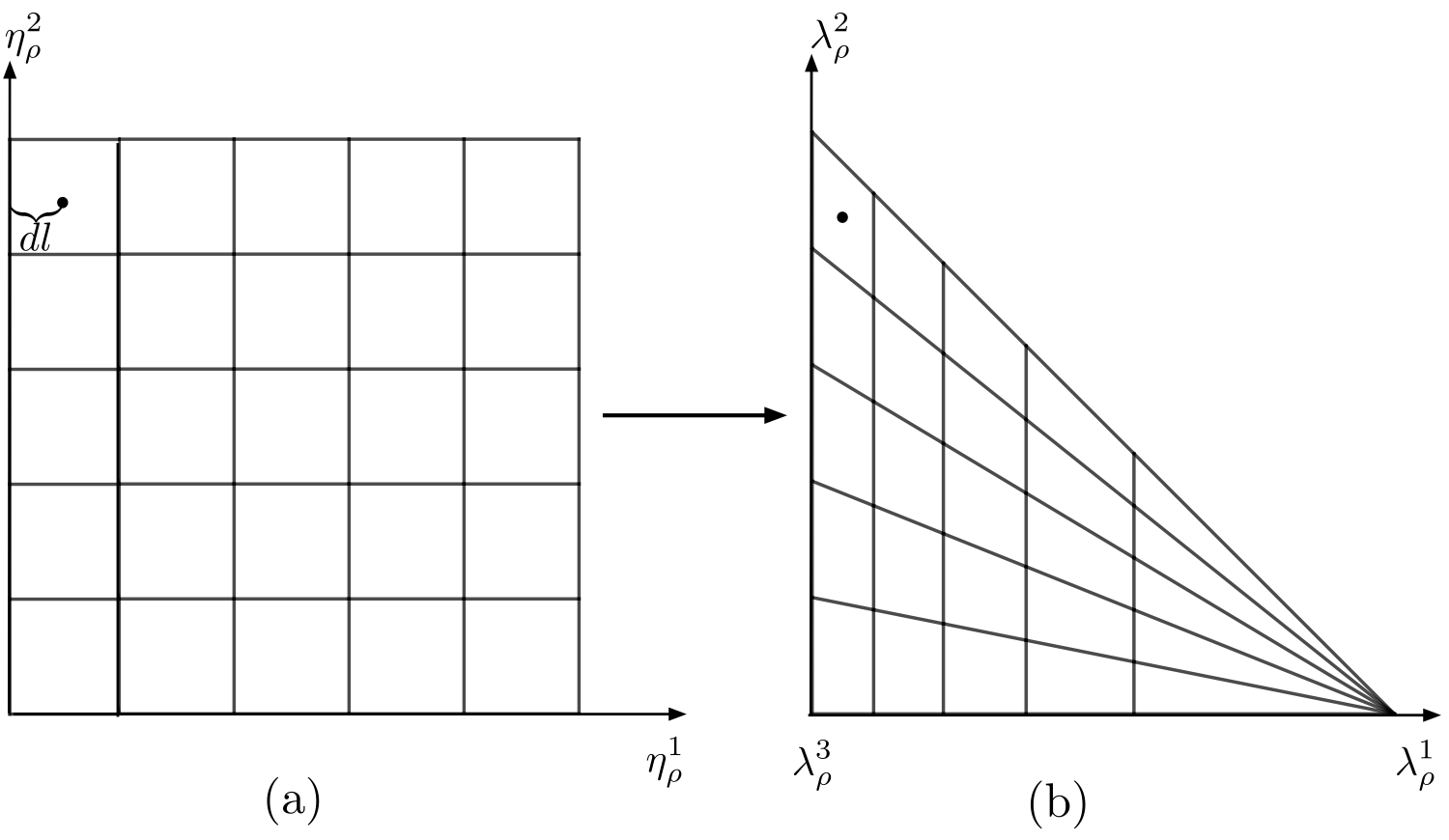}
\caption{Division of the probability simplex $\mathcal{S}$ into discrete $\rho_l$ of equal area for $\ell=5$.  In (a), we have the division of $\mathcal{S}$ in the $\vec{\eta}_{\rho}$ basis while (b) is in the $\vec{\lambda}_{\rho}$ basis.  The transformation is given by Eqs.~\ref{eq:lamb1} - \ref{eq:lamb3}. }
\label{fig:grid}
\end{figure}

Recall from Sec.~\ref{sec:back}, that the coarse-graining of $\gamma$-space is characterized by index sets $J_a \subset J$ that contain indices of combinatorially equivalent occupation sets $Z_j$'s.  Since Boltzmann's coarse-graining uses a finite number of particles and a finite discretization of $\mu$-space, the cardinality $|\{Z_j\}_j|=Z$ is finite.  Therefore, one can compute the fraction of $\gamma$-space $|J_a|/Z$ belonging to each macro-state.  To compute fractional volumes of $\mathcal{M}$ belonging to $\mathcal{M}_{\rho}$ that are comparable to $|J_a|/Z$, we must discretize $\mathcal{S}$ and $V^{\textrm{norm}}_{SO(3)} \in [0,1]$.

Ideally, one would compute $V_{\mathcal{M}_{\rho}}/V_{\mathcal{M}}$ and show that majority of $\mathcal{M}$ is subsumed by volumes associated with $\rho$'s with high information entropy.  But since $\rho$ is defined by its eigenvalues, which are continuous, it is not possible to define an exact density operator.  Therefore, one must divide $\mathcal{S}$ into discrete density operators $\rho_l$ and determine which $\rho_l$ are equivalent so that a fractional volume can be computed. Once an equivalence relation is determined, one simply needs to count the number of $\rho_l$ belonging to each equivalence class and divide by the total number of $\rho_l$ to get the fractional volumes.  

\begin{figure*}[t]
    \centering
    \includegraphics[width=5.5in]{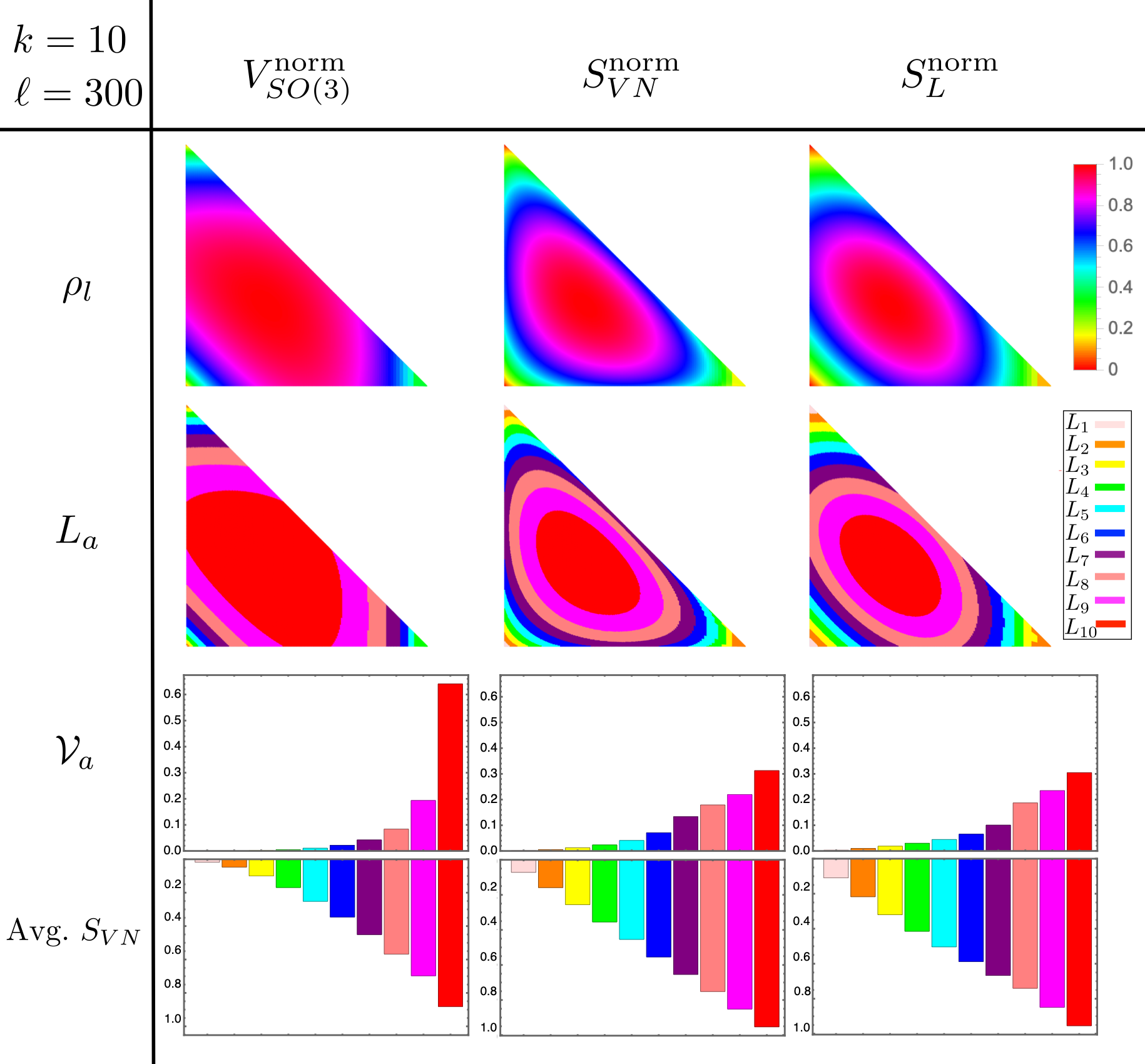}
    \caption{Results of coarse-graining $\mathcal{H}_{RA}=\mathbb{R}^3 \otimes \mathbb{R}^3$.  The first row, is the discretization of $\mathcal{S}$ where each $\rho_l$ is colored using the measure of each column. Row two is the result of discretizing the interval $[0,1]$ and sorting equivalent $\rho_l$ into segments $L_a$. Row three is the fraction of $\rho_l$ belonging to each $L_a$.  And row four is the average von Neumann entropy of each $L_a$.  It should be noted that the data from the graphs does not include the triangular distortions caused by the discretization of $\mathcal{S}$.  We only used data from Weyl chambers that do not include triangles.}
   \label{fig:results}
\end{figure*}

We discretize $\mathcal{S}$ into finite $\rho_l$ by uniformly sampling it using the transformation
\begin{eqnarray}
\label{eq:lamb1}
\lambda_{\rho}^1=1-\sqrt{\eta^1_{\rho}} \\
\label{eq:lamb2}
\lambda_{\rho}^2=\sqrt{\eta^1_{\rho}}(1-\eta^2_{\rho}) \\
\label{eq:lamb3}
\lambda_{\rho}^3=\sqrt{\eta^1_{\rho}}\eta^2_{\rho},
\end{eqnarray}
where $\eta^1_{\rho},\eta^2_{\rho} \in [0,1]$, as seen in~\cite{etas}. Dividing $\eta^1_{\rho}$ and $\eta^2_{\rho}$ into $\ell$ equal segments and transforming back to the $\vec{\lambda}_{\rho}$ basis divides $\mathcal{S}$ into $\ell^2$ discrete $\rho_l$, where $l\in[1,\ell^2]$; this is shown in Fig.~\ref{fig:grid}b.  To define our equivalence classes, we divide the segment $L=[0,1]$ into $k$ equal segments, $L_a$, where $a$ is an integer between $[1,k]$.  We treat $\rho_l$ whose volumes belong to a segment $L_a$ as equivalent.  Therefore, the fractional volumes are defined as
\begin{equation}
\label{eq:fracvol}
\mathcal{V}_a \equiv \frac{V_{\mathcal{M}_{\rho_a}}}{V_{\mathcal{M}}}=\frac{|L_a|}{|\rho_l|}
\end{equation} 
where $|L_a|$ is the number of $\rho_l$ belonging to $L_a$ and $|\rho_l|$ is the total number of discrete density operators. 

Since $V^{\textrm{norm}}_{SO(3)}$, $S^{\textrm{norm}}_L$, and $S^{\textrm{norm}}_{VN}$ are scalar functions on $\mathcal{S}$, they can be compared directly.  And since they all lie on the interval L, we can compute Eq.~\ref{eq:fracvol} for all three using the same procedure.  We simply replace volume with entropies when sorting $\rho_l$ into equivalence classes.

Choosing $\ell=300$ and $k=10$, we compute $V^{\textrm{norm}}_{SO(3)}$, $S^{\textrm{norm}}_L$ and $S^{\textrm{norm}}_{VN}$ at the center of squares in the $\vec{\eta}_{\rho}$ basis and assign that value to the corresponding $\rho_l$ in the $\vec{\lambda}_{\rho}$ basis. From Fig~\ref{fig:grid}a, we see that the distance from the center of a given square is given by $dl=1/(2\,\ell)$.  As $\ell$ goes to infinity, $dl$ goes to zero, and the volume/entropies associated with the $\rho_l$ in the $\vec{\lambda}_{\rho}$ basis becomes more representative of the actual value at the center.  

Coloring each $\rho_l$ using a colormap derived from the volume and entropies assigned to them gives the first row of Fig.~\ref{fig:results}.  We then assign an arbitrary color to each segment $L_a$ and color each $\rho_l$ in accordance to the $L_a$ in which they belong; this gives the second row of Fig.~\ref{fig:results}. There is nothing special about the choice of colors; they are only meant to distinguish $L_a$.  Computing Eq.~\ref{eq:fracvol}  and plotting the results gives the third row in Fig.~\ref{fig:results}. Due to the triangular distortions of $\mathcal{S}$ by the transformation from $\vec{\eta}_{\rho}$ to $\vec{\lambda}_{\rho}$, these plots are produced with the restriction that $ \eta^1_{\rho} \in (1/4,1]$ and $\eta^2_{\rho} \in (1/2,1]$.  This guarantees the data in the analysis is within a Weyl chamber~\cite{geometryq} that does not include the triangular distortions of the grid in the $\vec{\lambda}_{\rho}$ basis.   Finally, the fourth row of Fig.~\ref{fig:results} is the average von Neumann entropy of each $L_a$.

Analyzing the results, we see from the first row of Fig.~\ref{fig:results} that $V^{\textrm{norm}}_{SO(3)}$ behaves like an information entropy.  It upper bounds $S^{\textrm{norm}}_L$ and $S^{\textrm{norm}}_{VN}$, it's monotonic w.r.t purity, and it is concave.  Given these properties, we can see that the missing information in $\rho$ as measured by $S^{\textrm{norm}}_L$ and $S^{\textrm{norm}}_{VN}$ is quantified by the multiplicity of the composite system as measured by $V^{\textrm{norm}}_{SO(3)}$. Unlike the linear and von Neumann entropies, the top of $V^{\textrm{norm}}_{SO(3)}$ is flatter, and the drop in height is much steeper near pure states.  This is consistent with Boltzmann's original coarse-graining since the majority of the volume of $\mathcal{M}$ is made of high entropy states.  We can further see how $V^{\textrm{norm}}_{SO(3)}$ is consistent with Boltzmann's original coarse-graining by looking at rows three and four of Fig.~\ref{fig:results}.

The third and fourth rows show that over sixty percent of $\mathcal{M}$ consist of states with an average normalized von Neumann entropy of $0.88$ bits.  This is in stark contrast to the normalized von Neumann and linear entropies where only about thirty percent of $\mathcal{S}$ consist of states with an average von Neumann entropy above $0.88$ bits.  The drop off in the volume associated with pure states as seen in the first row is also made clear from the plots.  

In Boltzmann's analysis, over $99.99\%$ of $\gamma$-space consists of states at equilibrium.  This is because it is assumed that one is working with a high dimensional system with a number of particles on the order of Avagadro's number.  In this example, we are only working with $3$-level systems so the dimension of the space is vastly less.  Nonetheless, we still showed that the majority of phase space $\mathcal{M}$ is subsumed by states near equilibrium.  In Sec.~\ref{sec:son}, we show that our results tend toward the $99.99\%$ standard as the dimension of the system goes to infinity.

\subsection{Example: $SU(2)$}
\noindent Here, we give an example for a general qubit.  From Eq.~\ref{eq:UN} the unitaries of $SU(2)$ are given by
\begin{equation}
U(\phi,\psi,\chi)=
\begin{bmatrix}
e^{i \psi}\cos{\phi} && e^{i \chi} \sin{\phi}\\
-e^{-i \chi}\sin{\phi} && e^{-i \psi}\cos{\phi}
\end{bmatrix}
\end{equation}
where $\alpha=0$, $\psi, \chi \in [0,2\pi]$, $\phi \in [0,\pi/2]$, and the subscript $12$ 
\black{in the angles}
is dropped since the example is only 2-dimensional.  Computing the metric components directly, the nonzero values of the metric are
\begin{eqnarray}
g_{\phi \phi} &=& \lambda^1_{\rho} + \lambda^2_{\rho} \\
g_{\psi \psi} &=&  \left(\lambda^1_{\rho} + \lambda^2_{\rho}\right) \cos^2{\phi} \\
g_{\chi \chi} &=&  \left(\lambda^1_{\rho} + \lambda^2_{\rho}\right) \sin^2{\phi} \\
g_{\phi \psi} &=& g_{\phi \chi} = i(\lambda^1_{\rho}-\lambda^2_{\rho})\cos{\phi}\sin{\phi}. \\
\end{eqnarray}
Taking the $\sqrt{\hbox{Det}(\bold{g})}$ and integrating over $\{\phi,\psi,\chi\}$ gives
\begin{equation}
\label{eq:su2vol}
V_{SU(2)}=\sqrt{\lambda^1_{\rho}(1-\lambda^1_{\rho})}\; \black{= \sqrt{S_L/2}}
\end{equation}
where $\lambda^2_{\rho}=1-\lambda^1_{\rho} \black{= \tfrac{1}{2}\left[ 1+\sqrt{2\Tr[\rho^2]-1}\right]}$.

\begin{figure}[h]
\centering
\includegraphics[width=\columnwidth]{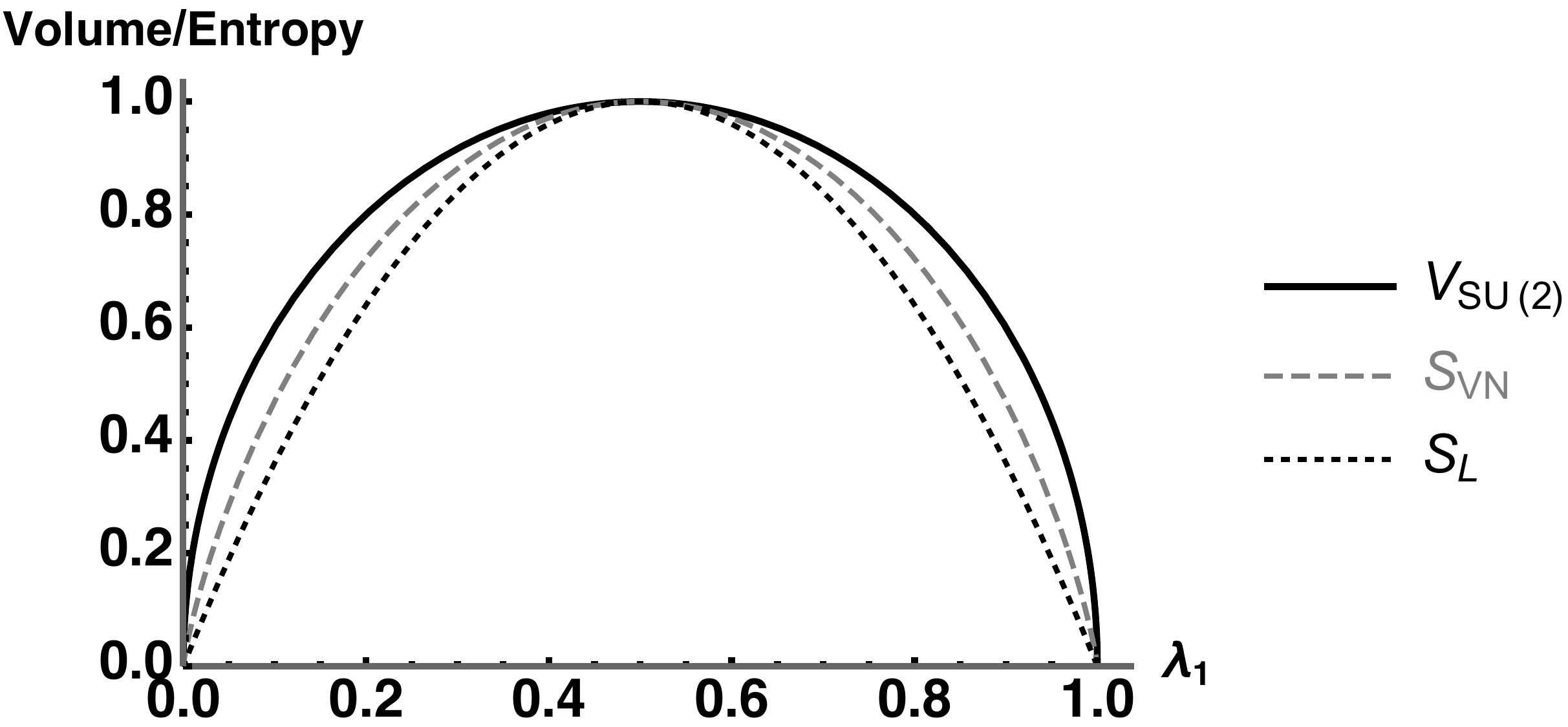}
\caption{
\black{Plot of} the normalized volume, linear and von Neumann entropies for 2-level systems whose symmetries are $SU(2)$.}
\label{fig:twod}
\end{figure}

From Eq.~\ref{eq:su2vol}, we again see that our volume has the desired traits of an information entropy.  It is maximal on a maximally mixed state, zero on pure states, and monotonic.  
\black{In fact, in this case, $2\,V^2_{SU(2)} = S_L = 1~-~\Tr[\rho^2]$, the linear entropy.}
\black{Fig.~\ref{fig:twod} compares the volume of $SU(2)$ with the linear and von Neumann entropies.}
Here we see that the volume has the same traits for $SU(2)$ as it did for $SO(3)$.  
Again, the volume 
\black{is directly related}
to \black{both} the linear and von Neumann entropies.
It bounds both of them from above; 
it is flatter in the region near maximally mixed states; it is concave w.r.t $\lambda^1_{\rho}$,
and it has a steeper drop off as the states become more pure.  This shows again that the missing information in $\rho$ as measured by $S^{\textrm{norm}}_L$ and $S^{\textrm{norm}}_{VN}$ is quantified by $V^{\textrm{norm}}_{SO(3)}$.

\subsection{Example: $SO(N)$}
\label{sec:son}
\noindent We complete our examples of volumes with a discussion on $SO(N)$.  If we set $\psi=\xi=0$ for our 2D example, we are left with one non-zero metric component $g_{\phi \phi}=\sqrt{\lambda^1_{\rho}+\lambda^2_{\rho}}=1$.  As we see from Eq.~\ref{eq:vola}, the volume is merely the square root of the product of all pairwise sums of eigenvalues.  Using the same procedure to compute the volume for $SU(2)$ and $SO(3)$, we computed the volume of $SO(4)$ 
and \black{obtained}
\begin{equation}
\label{eq:so4}
V_{SO(4)}=\prod^4_{i<j}{\sqrt{\lambda^i_{\rho}+\lambda^j_{\rho}}}.
\end{equation}
%
\black{By induction, we infer that the volume for $N$-level systems with no complex phases is given by}
\begin{equation}
\label{eq:soN}
V_{SO(N)}=\prod^N_{i<j}{\sqrt{\lambda^i_{\rho}+\lambda^j_{\rho}}}.
\end{equation}
Using this equation of volume for arbitrary dimension, we show that the percentage of phase space subsumed by states with high entropy increases as $N\rightarrow \infty$.

\black{Classically, it} is known that for high dimensional systems with many particles, over \black{$99.99\%$} of phase space consists of configurations with states near or at equilibrium~\cite{goldstein,goldplus,tasaki}. So far, we have shown in Fig.~\ref{fig:results} that higher entropy states take up the majority of phase space for $3$-level systems.  If our coarse-graining is consistent with Boltzmann, one would expect the percentage of phase space consumed by states near or at equilibrium to tend toward \black{$99.99\%$}. 

\begin{figure}[h]
\centering
\includegraphics[width=\columnwidth]{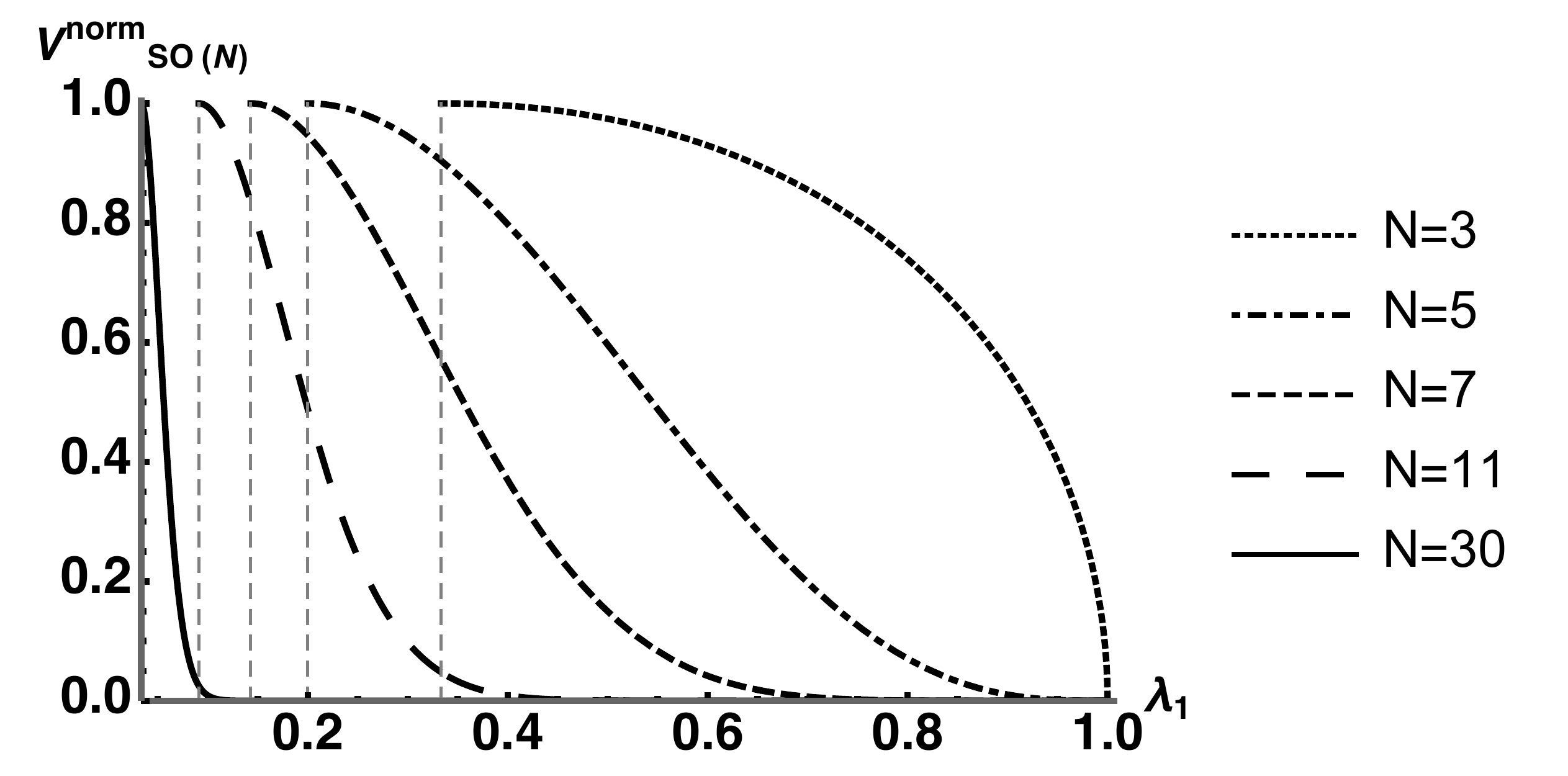}
\caption{
\black{Plot of} $V^{\textrm{norm}}_{SO(N)}$ for $N=3,5,7,11,30$.  The dashed verticle lines are located at the minimal value of $\lambda^1_{\rho}$ for each plot, which is $1/N$, \black{the maximally mixed state}.  Notice how the centroids tend toward maximally mixed states as pure states subsume less volume as $N$ increases.}
\label{fig:volnplots}
\end{figure}

\begin{figure}[h]
\centering
\includegraphics[width=2.7in]{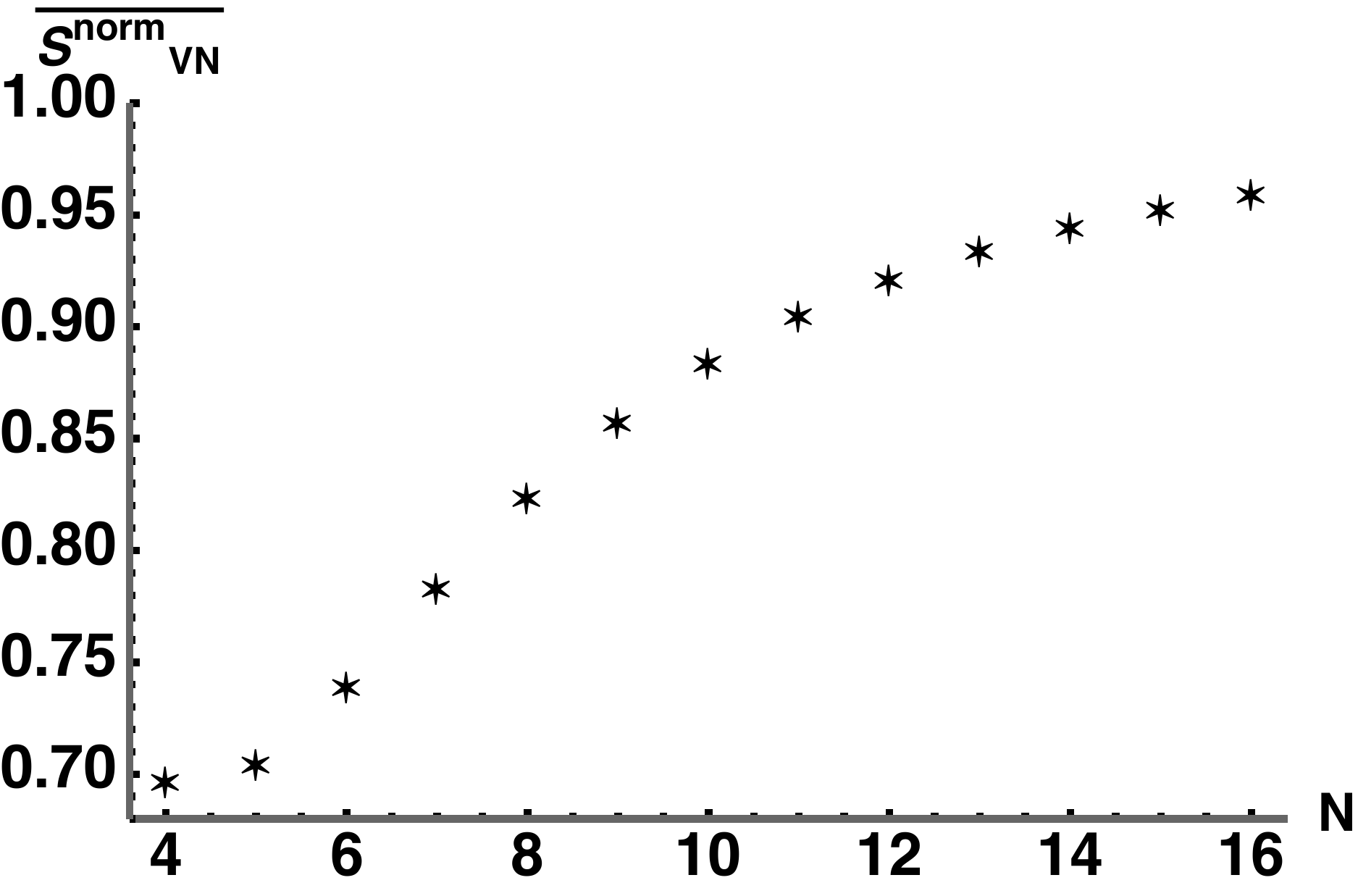}
\caption{
\black{Plot of} the average von Neumann entropy 
\black{(normalized to the maximally mixed state)}
between $\lambda_{\rho}^1 \in [1/N,\lambda_{\rho}^{1*}]$ as a function of N.  This quantifies the results of Fig.~\ref{fig:volnplots} by showing that the average von Neumann entropy of states whose volumes take over 
\black{$99.99\%$} 
of phase space increases.}
\label{fig:avgsvn}
\end{figure}

In order to show high-entropy states comprise most of the volume of phase space, we consider the set of marginal density matrices that are mixtures of a pure state and the maximally mixed state whose mixedness is parametrized by its first eigenvlaue $\lambda_{\rho}^{1}$. Letting $\lambda^i_{\rho}=(1-\lambda^1_{\rho})/(N-1)$ for all $i\in\{2,N\}$, we can write \Eq{eq:soN} as a function of 
$\lambda^1_{\rho}$. Inserting this choice of eigenvalues and normalizing w.r.t the maximum volume gives
\begin{equation}
\label{eq:volnlamb1}
V^{\textrm{norm}}_{SO(N)}=\frac{\left(\lambda^1_{\rho}+ \frac{1-\lambda^1_{\rho}}{N-1}\right)^{\frac{N-1}{2}}\left( 2\frac{1-\lambda^1_{\rho}}{N-1}\right)^{\frac{(N-1)(N-2)}{4}}}{\left(\frac{2}{N}\right)^{\frac{N(N-1)}{4}}}.
\end{equation}
To show that the volume of phase space increasingly tends toward maximally mixed states, we plot Eq.~\ref{eq:volnlamb1} for $N=3,5,7,11,$ and $30$ in Fig.~\ref{fig:volnplots}.
In this figure, we see that the centroid of each plot tends toward maximally mixed states as $N$ increases.  To quantify these results, we identify the value $\lambda_{\rho}^{1*}$ for \black{various} values of $N$ where $V^{\textrm{norm}}_{SO(N)}(\lambda_{\rho}^{1*})=10^{-4}$.  For the values of $N$ used, this choice of $\lambda_{\rho}^{1*}$ guarantees that 
\begin{equation}
\label{eq:lambstar}
\frac{\int_{1/N}^{\lambda_{\rho}^{1*}}{V^{\textrm{norm}}_{SO(N)}(\lambda_{\rho}^{1})\ d\lambda^1_{\rho}}}{\int_{1/N}^{1}{V^{\textrm{norm}}_{SO(N)}(\black{\lambda_{\rho}^{1}})\ d\lambda^1_{\rho}}} > 0.9999,
\end{equation}
\black{where $\lambda_{\rho}^{1}= 1/N$ indicates the maximally mixed state.}
Plotting the average normalized 
\black{(to the maximally mixed state)}
 von Neumann entropy between $\lambda_{\rho}^1 \in [1/N,\lambda_{\rho}^{1*}]$ as a function of N gives Fig.~\ref{fig:avgsvn}.  This clearly shows that the majority of the phase space volume tends toward maximally mixed states \black{with increasing $N$}.

From this analysis, we have shown that our canonical quantum coarse-graining reproduces the fractional volume distribution of macro-states in Boltzmann's original coarse-graining for the examples considered.  We also showed for $SO(3)$ and $SU(2)$ that the missing information in $\rho$ as measured by information entropies is directly related to the multiplicity of the composite system.  This is consistent with Brillouin's approach for relating information and thermal entropies.  We did not include an analysis of $SU(N)$ since computing the determinant becomes prohibitively difficult as the number of 
phase space parameters increases~\black{\cite{SUN:comment}}.

\section{Flag Varieties and Qualitative Physical Examples}
\label{sec:flags}
\noindent The purpose of this section is to provide conceptual examples of 
\black{our} 
 canonical quantum coarse-graining 
\black{that we have introduced}.  To \black{this end}, we focus on two topics.  First, we discuss the inherent 
flag variety structure~\cite{hilbertflag} that underlies our coarse-graining.  This is meant to help the reader see how the structure of the coarse-graining is well defined and unique for any state.  \black{Subsequently}, we present some physical examples to further understand the surfaces of ignorance and their volumes in less abstract terms.


\subsection{Flag Varieties}
\label{sec:flags}
\noindent We begin by discussing the features of flag varieties that inherently arise in our coarse-graining.  As structures, flag varieties function as a means of resolving manifolds with greater resolution.  This is because flag varieties are defined as a sequence of nested submanifolds
\begin{equation}
\label{eq:flag}
F_M=V_1 \subset V_2 \subset . . . \subset V_{m-1} \subset V_m \subset V_{m+1}\subset ... \subset V_M
\end{equation}
where $dim(V_n) < dim(V_m)$ for $n<m$.  Each submanifold $V_{m+1}$ resolves points in $V_m$ with greater resolution.  This is analogous to using a microscope on a point in $V_m$ to reveal greater detail that could not be resolved before.  In the case of 
\black{our}
canonical quantum coarse-graining, $\mathcal{H}_A$ is analogous to $V_m$ and $\mathcal{H}_R\otimes \mathcal{H}_A$ is analogous to $V_{m+1}$.  The hallmark of coarse-graining is the process of moving between high-resolution data to low-resolution data.  From Eq.~\ref{eq:flag}, it should be clear how this property is inherently captured by flag varieties.  The next defining property of flag varieties that we discuss is their construction through transitive group actions.  This trait ensures that $\mathcal{H}_{RA}$ (higher \black{dimensional} Hilbert space) of $\mathcal{H}_A$ (lower \black{dimensional} Hilbert space) is uniquely demarcated into mutually exclusive sets, which is another hallmark of coarse-graining. 


The general relationship between micro- and macro-states is captured by the property of transitive group actions, which is another defining characteristic of flag varieties.  A group action on a set $X$ is transitive when $X$ has exactly one orbit associated with the group.  This means for a single point (density operator) in a lower \black{dimensional} Hilbert space, the set of purifications in the higher Hilbert space is generated by a single group action.  This is visualized in Fig.~\ref{fig:fib2} where $SO(3)$ acting on $\rho$ generates $\{\bar{\Gamma}^{\rho}(\vec{\xi})\}$.  Here, $X \equiv \{\bar{\Gamma}^{\rho}(\vec{\xi})\}$ and is completely constructed from the single orbit of $\rho$.  This feature ensures that all macro-states defined as points in a lower  \black{dimensional} Hilbert space are uniquely identified by a single set of purifications in a higher  \black{dimensional} Hilbert space.  With the transitive property understood, we can now discuss how one can access all points in the entire flag structure through points in all subspaces.  

\begin{figure}[h]
\centering
\includegraphics[width=\columnwidth]{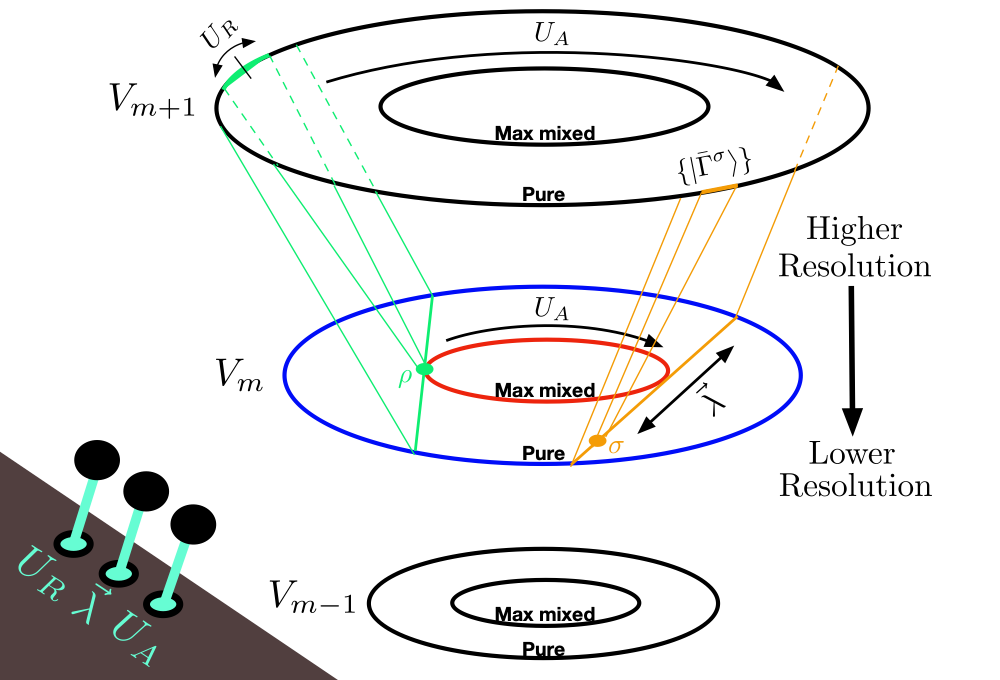}
\caption{
\black{Conceptual illustration of} how points in the higher dimensional subspace can be defined by points in lower dimensional subspaces.  The levers in the foreground represent the freedom to change settings to uniquely identify a point in the higher resolution space.  The $U_R$ stick cycles through unitaries in the reservoir space which is equivalent to cycling through purifications of density operators $\rho$ and $\sigma$ defined in the lower  \black{dimensional} space.  The $\vec{\lambda}$ stick cycles through settings of eigenvalues, and the $U_A$ stick cycles through unitaries in the base space. This also cycles through pure states in the higher space since each simplex in the base space has a set of pure states in the higher  \black{dimensional} space associated with it.}
\label{fig:flag}
\end{figure}

To see how the flag variety functions indefinitely, we refer to Fig.~\ref{fig:flag}.  The dimension of the subspaces increase from the bottom to the top.  To understand the figure, assume that we are working in a perspective where all density operators are diagonalized as in Fig.~\ref{fig:tor3}.  For reference, imagine our $SO(3)$ example where $V_m=\mathbb{R}^3$ and $V_{m+1}=\mathbb{R}^3\otimes \mathbb{R}^3$.  Consider that each point on the flag structure can be accessed through settings of eigenvalues and unitary transformations, and each lever in Fig.~\ref{fig:flag} represents the freedom to change the settings.  The $U_A$ lever applies unitary transformations in the base space.  That is, one can rotate the orange simplex into the green simplex using this stick.  The $\vec{\lambda}$ lever allows one to slide the point on the simplex.  This is equivalent to choosing eigenvalues and thus defining the density operator in the base space. The $U_R$ lever allows one to change the settings of the phase in the higher  \black{dimensional} Hilbert space.  This is equivalent to choosing $\vec{\xi}$ in $\{\bar{\Gamma}(\vec{\xi})\}$ which chooses a specific purification and fixes the gauge. With these three levers, one can choose any pure state in the higher space by selecting a point in the lower space.  They can then select any point by taking mixtures of these pure states.  For each higher  \black{dimensional} Hilbert space, a mixture of pure states can be constructed whose purifications exist in an even higher \black{dimensional} Hilbert space.  With the flag variety structure, this process is always well defined and can be continued ad infinitum.  

\black{Having} discussed the inherent flag variety structure that emerges from 
\black{our}
canonical quantum coarse-graining, 
\black{we turn now to some}
qualitative physical interpretations of the surfaces of ignorance and the space of speculation.  
This is designed to \black{provide greater}  intuition about the coarse-graining.  Rigorous considerations of the examples given are left for future investigations. 

\subsection{Qualitative Physical Examples}
\label{sec:physexamp}
\noindent For our first example, \black{the process of} thermalization as seen from an observer whose information is limited to  lower-resolution data in   \black{a lower dimensional} Hilbert space, and an observer with access to higher-resolution data in  \black{a higher dimensional} Hilbert space.  Here, a \black{\textit{thermalization process}} is merely \black{some transition} where a state goes from higher to lower purity due to interactions with its environment.   

\begin{figure}[h]
\centering
\includegraphics[width=3.2in]{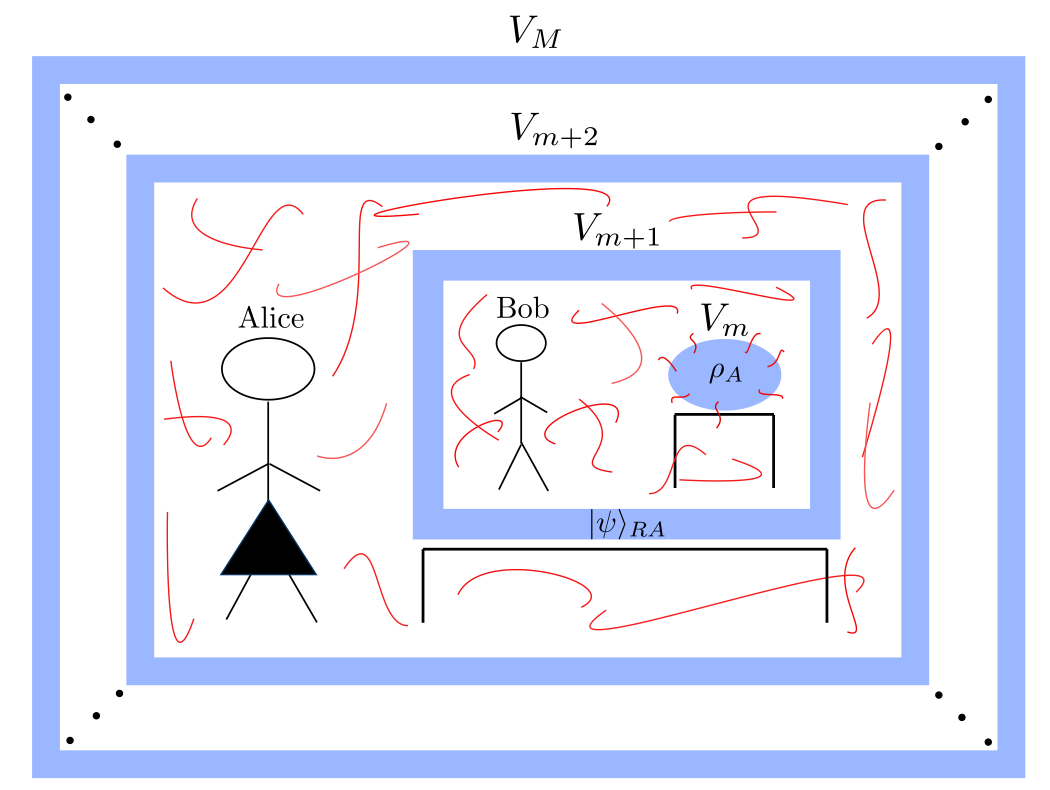}
\caption{
\black{Conceptual illustration of}
 a thermalization process between $\rho_A \in V_m$ \black{and its environment $V_{m+1}$}.  
 The state $\rho_A$ is thermalized from interactions with the surrounding environment. For Bob, this evolution is not unitary.  Alice is a global observer and thus has full information of $\rho_A$ and the environment as given by $|\psi \rangle_{RA} \in V_{m+1}$.  For Alice, $|\psi \rangle_{RA}$ is sufficiently isolated from her environment and thus thermalization of $\rho_A$ is a unitary evolution of $|\psi \rangle_{RA}$.  This thermalization example can be extended ad infinitum continuously embedding states into higher dimensional Hilbert spaces.}
\label{fig:comptherm}
\end{figure}

In Fig.~\ref{fig:comptherm}, we give a physical example of a thermalization process in the context of a flag variety structure.  Bob is treated as a local observer within the composite state $|\psi \rangle_{RA} \in V_{m+1}$, which consists of the subsystem $\rho_A \in V_m$ and a noisy environment represented by the red squiggly lines.  As a local observer, Bob is incapable of having full knowledge about $|\psi \rangle_{RA}$.  Instead, his knowledge is restricted to the information in $\rho_A$.  From Bob's perspective, any interaction between $\rho_A$ and the environment results in a non-unitary evolution given by
\begin{equation}
\label{eq:evobob}
\rho_A(t)=\Tr_{\black{R}}[U_{RA}(t)|\psi \rangle_{RA}\langle\psi|U^{\dagger}_{RA}(t)],
\end{equation}
which 
\black{provides the mechanism for his increasing ignorance} about the state of $\rho_A$.  
These interactions are represented by the flux of red lines emanating from $\rho_A$.  

Unlike Bob, Alice 
is treated as a global observer outside of the composite system
\black{of Bob and his environment}.  
This gives her access to full information of $|\psi \rangle_{RA}$.  For her, $|\psi \rangle_{RA}$ is a pure state \black{embeded within an even} larger environment.  But unlike $\rho_A$, there is no flux between $|\psi \rangle_{RA}$ and the \black{larger} environment in Alice's lab.  This is shown in Fig.~\ref{fig:comptherm} by a lack of flux between $|\psi \rangle_{RA}$ and the environment of Alice's lab. Therefore, from Alice's perspective, the thermalization of $\rho_A$ is a unitary evolution of $|\psi \rangle_{RA}$ given by
\begin{equation}
\label{eq:evoalice}
|\psi(t)\rangle_{RA}=U_{RA}(t)|\psi(t_0)\rangle_{RA}.
\end{equation}

Relating Fig.~\ref{fig:comptherm} to Fig.~\ref{fig:flag}, the thermal evolution as seen by Bob is given by the point on the green simplex sliding from the outer blue circle (representing pure states) of $V_m$ \black{towards} the inner red circle (representing maximally mixed states).  For Alice, this evolution would be along the outer pure state circle of $V_{m+1}$.  If there was flux between $|\psi \rangle_{RA}$ and Alice's environment, the evolution would fall inward from the pure state circle to the inner maximally mixed circle. The resulting state could be purified in $V_{m+2}$ and this process can be repeated indefinitely.  

To further examine our thermalization example, we \black{now examine} how these evolutions are observed by Alice and Bob in their respective spaces. \black{To this end}, we visualize the \black{higher dimensional} Hilbert space $V_{m+1}$ analogously to $\gamma$-space in Fig.~\ref{fig:mugamma}.  We also treat the lower  \black{dimensional} Hilbert space $V_{m}$ in which Bob lives as the base space seen in Fig.~\ref{fig:fib2}.  That is, \black{without loss of generality} we treat $\rho_A$ as a 3-level system with zero complex \black{phases}.  This is only for \black{the purpose} of visualization \black{in order to} illustrate the thermalization perspectives of different resolution.

\begin{figure}[h]
\centering
\includegraphics[width=2.5in]{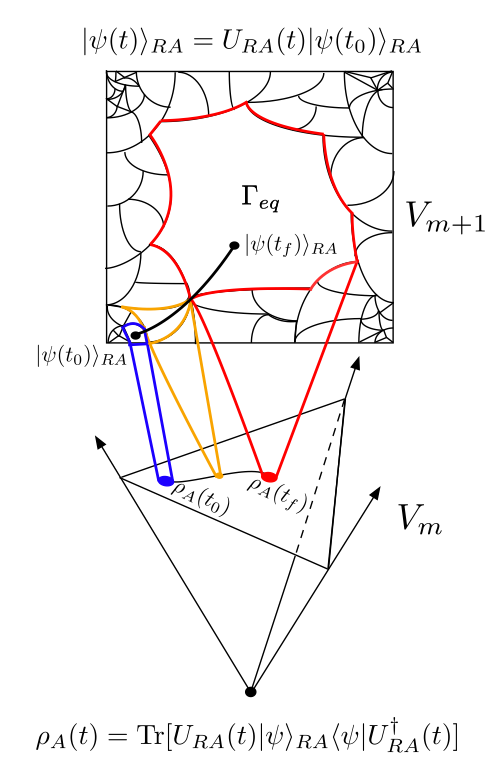}
\caption{
\black{Illustration of} the thermalization process from \black{both} Bob and Alice's perspectives.  Here, $U_{RA}(t)$ is the unitary describing the evolution of $|\psi \rangle_{RA}$.  Bob's knowledge is restricted to the lower space $V_m$ and thus the resolution of his model is limited by the dimension of that space.  Therefore, when looking from Bob's perspective in $V_{m+1}$, we are limited to sets of purifications that are consistent with $\rho_A(t)$ at each point of the evolution. This is due to the principle of insufficient reason and is depicted by the blue, orange, and red sets in the figure.  For Alice, her knowledge is restricted by the resolution of the higher space $V_{m+1}$.  If she knows the initial pure state, her knowledge of the thermalization process is a continuous curve of pure states from $|\psi(t_0)\rangle_{RA}$ to $|\psi(t_f)\rangle_{RA}$.  This is the black curve in $V_{m+1}$.}
\label{fig:thermalize}
\end{figure}

To see the thermalization processes from \black{both} Alice and Bob's perspectives, we refer to  Fig.~\ref{fig:thermalize}.  The upper portion of this figure represents the set of pure states associated with the probability simplex in the base space.  This is equivalent to the portion of $V_{m+1}$ in Fig.~\ref{fig:flag} that is bounded by the end points of the simplicies in $V_m$ of the same figure.  Since the only information available to Bob is in $\rho_A$, his knowledge of $|\psi \rangle_{RA}$ in the higher  \black{dimensional} Hilbert space can only be narrowed down to the set of purifications $\{\bar{\Gamma}^{\rho}\}$.   Therefore, in the higher  \black{dimensional} Hilbert space, Bob only sees the system evolve from sets of pure states to other sets of pure states \black{represented as closed regions bounded by colored perimeters}. 
 \black{This is illustrated} in Fig.~\ref{fig:thermalize} where the state begins in the small  set of purifications \black{bounded in blue}, has an intermediate step given \black{by the larger orange-bounded set}, and finally thermalizes to the largest 
 \black{red-bounded} set.  This \black{coarsely resolved evolution} is in \black{distinct} contrast to how Alice \black{observes} the evolution.

Since \black{Alice's} knowledge of the system is in the higher  \black{dimensional} Hilbert space, and there is no interaction between $|\psi \rangle_{RA}$ and her environment, all of her states during the evolution are pure.  That is, the information contained in $|\psi(t)\rangle_{RA}$ is always $\log{d_{RA}}$.  Because she always has full information, if she knows the initial state of the evolution, then she knows ``exactly" the trajectory taken during the thermalization process.  This is \black{denoted as} the solid black line in Fig.~\ref{fig:thermalize} whose end points are $|\psi(t_0)\rangle_{RA}$ and $|\psi(t_f)\rangle_{RA}$.

In the \black{above} thermalization example, we see how an evolution appears depending on the space in which it is observed.  For observers in the lower  \black{dimensional} Hilbert space, a thermalization process appears as a non-unitary evolution with increasing ignorance as captured by the volume of purifications.  For an observer in the higher  \black{dimensional} Hilbert space, the same thermalization process in a unitary evolution with no information loss.  From this example, it is easy to see how the coarse-graining captures ignorance for an indefinite number of observers in nested Hilbert spaces of increasing dimension.  As another example of the surfaces of ignorance, we \black{next} examine \black{how} a unitary transformation \black{appears in this context} in the lower  \black{dimensional} Hilbert space. 

\begin{figure}[h]
\centering
\includegraphics[width=3.5in]{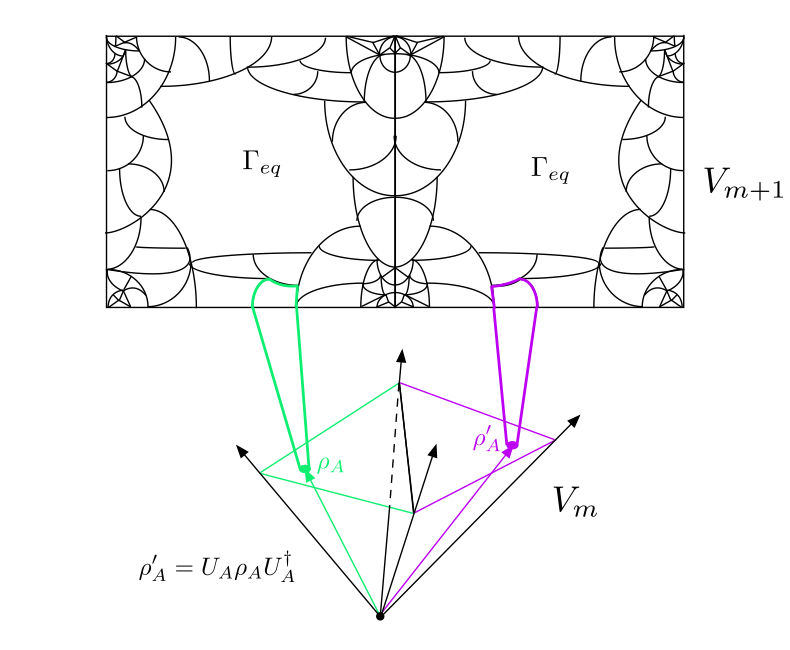}
\caption{
\black{Depiction of} a unitary transformation by Bob of $\rho_A$ in the base space.  For each choice of phase angles $\vec{\xi}_A$ of $U_A(\vec{\xi}_A)$ when transforming between probability simplicies, there exists a copy of the set of purifications for each simplex.  Since the eigenvalues of $\rho_A$ are not changed, the volume of the surface of ignorance in $V_{m+1}$ must remain fixed as shown in the figure.}
\label{fig:unitary}
\end{figure}
Assuming that $\rho_A$ is isolated from the environment and only evolves unitarily due to Bob's manipulations, we have Fig.~\ref{fig:unitary}.  
Here, a unitary has been applied to eigenbases $\{|\lambda^{i}_\rho \rangle\}$ \black{which passively rotates the}
 eigenvalue vector $\vec{\lambda}_{\rho}$ that simply flips the simplex.  
This results in another \black{``copy"} of the purifications associated with the simplex.  It is clear that the landscape of pure states in $V_{m+1}$ associated with all unitarily equivalent probability simplexes in $V_{m}$ must be equivalent.  Otherwise, it would imply that each simplex related by $U_A$ is distinguishable which implies that symmetries are not preserved during unitary transformation; this is obviously false.  Rather, each simplex only becomes distinguishable when a fixed simplex is used as a reference frame to compare all other simplicies. This is also true for the set of purifications associated with simplicies in $V_{m+1}$.  They are identical copies, but they differ with respect to a choice of phase in $V_m$.  This also makes clear why the volume of the surface of ignorance is fixed with unitary transformations of $\rho_A$. 
\black{Equivalently, $\rho_A$ and  $\rho'_A = U_A\,\rho_A\,U^\dagger_A$ have the same set of eigenvalues, and hence the same volume of the fibers above them, i.e. $V_{\rho_A} = V_{\rho'_A}$. This can be considered as the active-transformation point of view of the evolution $U_A$ transforming $F_{\rho_A}\to F_{\rho'_A}$~\cite{passact:comment}.}

\section{Conclusion}
\label{sec:conclusion}
\noindent In this paper we defined a canonical quantum coarse-graining using Lie group symmetries that are inherent to the Hilbert spaces in which density operators are defined.  Given this choice of symmetries, macro-states are defined as density operators in the base space $H_A$, and micro-states are elements of the fiber/set of purifcations $F_{\rho}=|\bar{\Gamma}^{\rho}(\vec{\xi})\rangle$ of $\rho$.  The fiber is a subspace of the higher \black{dimensional} Hilbert space $H_{RA}\equiv \mathcal{H}_R \otimes \mathcal{H}_A$ and is unique to $\rho$.  
Therefore, they are also defined as macro-states .
%
Since purifications in $\mathcal{H}_{RA}$ \black{are} the space in which all possible configurations of the system are represented given the constraint data $\rho$, 
\black{$\mathcal{H}_{RA}$ can be considered the phase space of $\mathcal{H}_A$ using the broadest possible definition of a phase space}.
This guarantees a demarcation of phase space into disjoint sets by the definition of purifications.  

\black{The elements of $F_{\rho}$ are generated using representations of Lie group symmetries, which are merely unitary transformations of $\mathcal{H}_R$ and $\mathcal{H}_A$.  The metric components of the differential manifold $\mathcal{M}_{\rho}$ are constructed from $\rho$ and its accompanying fiber $F_\rho$, and each point on $\mathcal{M}_{\rho}$ is a purification of 
$\rho$}.  
We defined these manifolds as surfaces of ignorance and showed that their volumes are related to the amount of information missing in $\rho$ as measured by information entropies.  To demonstrate that these volumes function as an information entropy, and to show that our coarse-graining reproduces the properties of Boltzmann's original coarse-graining, we analyzed the volumes for states whose symmetries are defined by $SO(3)$, $SU(2)$, and $SO(N)$.   

The consequences of our coarse-graining and the results of our analysis are the following.  (1) We showed that our volume functions as an information entropy by comparing it to the linear and von Neumann entropies.  In fact, we showed that it is an upper bound for both.  (2) Our coarse-graining reproduces the relationship between quantum information and thermal entropies that was shown for classical information and thermal entropies by the analysis of Brillouin~\cite{brillouin}.  In particular, by showing that our volume \black{upper bounds the} information entropies, 
we demonstrated that an increase of negentropy is consistent with the decrease in possible micro-configurations of the system.  (3) The resulting demarcation of phase space is consistent with the properties of Boltzmann's original coarse-graining where the majority of phase space is subsumed by states near or at equilibrium .  This was demonstrated in \Sec{sec:volana} using a detailed coarse-graining of states defined by $SO(3)$ symmetries. We also showed this for $SO(N)$ as \black{demonstrated} in Figs.~\ref{fig:volnplots} and~\ref{fig:avgsvn}. (4) A general formula of volume, Eq.~\ref{eq:soN}, for $SO(N)$ was given.  (5)  For any state $\rho$, there is one unique coarse-graining given by our procedure.  This coincides with the inherent flag variety structure that emerges, and guarantees a well defined coarse-graining of composite Hilbert spaces ad infinitum. (6)  This procedure connects concepts \black{such as} fiber bundles, gauge theories, data compression, thermalization, and quantum information and thermal entropies using coarse-graining defined by Lie group symmetries of Hilbert spaces. 

The claim that this procedure is 
\black{a canonical one}
is based on the fact that the coarse-graining is unique given a density operator 
$\rho$.  
\black{As far as we are aware, this cannot be said for any other proposed quantum coarse-graining procedures}.  
\black{The use of term canonical is also appropriate since we have defined our coarse-graining procedure by}
using symmetries and definitions of micro- and macro-states inherent to any Hilbert space.  \black{Our} procedure is always uniquely defined and the structure of the coarse-graining always exists regardless of the states considered.  For future research, we plan to formalize a notion of quantum Boltzmann entropy using this coarse-graining \black{method}. 

\begin{acknowledgments}
The authors wish to thank Christopher C. Tison and James Schneeloch for many useful discussions and inputs.  PMA would like to acknowledge support of this work from
the Air Force Office of Scientific Research (AFOSR).
CC is grateful to the United States Air Force Research Laboratory (AFRL) Summer Faculty Fellowship Program for providing support for this work
under grant \#FA8750-20-3-1003.
Any opinions, findings and conclusions or recommendations
expressed in this material are those of the author(s) and do not
necessarily reflect the views of Air Force Research Laboratory. 
\end{acknowledgments}
\clearpage
\newpage
\nocite{apsrev41Control}
\bibliographystyle{apsrev4-1}
\bibliography{cqcg_redone_3Oct2021} 

\end{document}